\documentclass{aa}

\usepackage{natbib}
\bibpunct{(}{)}{;}{a}{}{,} 

\usepackage{graphicx}
\usepackage{txfonts}
\usepackage{xcolor}
\usepackage[colorlinks=true,
    citecolor=blue,linkcolor=blue]{hyperref}
\begin{document}

   \title{Imaging of I Zw 18 by JWST: II. Spatially resolved star formation history}


   \author{Giacomo Bortolini \inst{1}
          \thanks{Email: giacomo.bortolini@astro.su.se}
          \and
          G{\"o}ran {\"O}stlin\inst{1}
          \and
          Nolan Habel \inst{2}
          \and
          Alec S. Hirschauer\inst{3}
          \and
          Olivia C. Jones \inst{4}
          \and 
          Kay Justtanont \inst{5}
          \and
          Margaret Meixner \inst{2}
          \and
          Martha L.\ Boyer \inst{3}
          \and
          Joris A.\ D.\ L.\ Blommaert \inst{7}
          \and 
          Nicolas Crouzet \inst{8}
          \and
          Laura Lenki\'{c} \inst{6,2}
          \and
          Conor Nally \inst{9}
          \and 
          Beth\ A.\ Sargent \inst{3,10}
          \and
          Paul van der Werf \inst{8}
          \and
          Manuel G\"udel \inst{11,12}
          \and
          Thomas Henning\inst{13}
          \and
          Pierre \ O. Lagage \inst{14}
          }

   \institute{
    Department of Astronomy, The Oskar Klein Centre, Stockholm University, AlbaNova, SE-10691 Stockholm, Sweden
    \and
    Jet Propulsion Laboratory, California Institute of Technology, 4800 Oak Grove Dr., Pasadena, CA 91109, USA
    \and
    Space Telescope Science Institute, 3700 San Martin Drive, Baltimore, MD 21218, USA
    \and
    UK Astronomy Technology Centre, Royal Observatory, Blackford Hill, Edinburgh, EH9 3HJ, UK
    \and
    Chalmers University of Technology, Dept.\ Space, Earth and Environment, Onsala Space Observatory, 439 92 Onsala, Sweden
    \and
    Stratospheric Observatory for Infrared Astronomy, NASA Ames Research Center, Mail Stop 204-14, Moffett Field, CA 94035, USA
    \and
    Astronomy and Astrophysics Research Group, Dep.\ of Physics and Astrophysics, V.U.\ Brussel, Pleinlaan 2, B-1050 Brussels, Belgium
    \and
    Leiden Observatory, Leiden University, P.O.\ Box 9513, 2300 RA Leiden, The Netherlands
    \and
    Institute for Astronomy, University of Edinburgh, Blackford Hill, Edinburgh, EH9 3HJ, UK
    \and
    Department of Physics \& Astronomy, Johns Hopkins University, 3400 N.\ Charles St., Baltimore, MD 21218, USA
    \and
    Dept. of Astrophysics, University of Vienna, T\"urkenschanzstr. 17, A-1180 Vienna, Austria
    \and
    ETH Z\"urich, Institute for Particle Physics and Astrophysics, Wolfgang-Pauli-Str. 27, 8093 Z\"urich, Switzerland
    \and
    Max-Planck-Institut f\"ur Astronomie (MPIA), K\"onigstuhl 17, 69117 Heidelberg, Germany
    \and 
    Universit\'e Paris-Saclay, Universit\'e Paris Cit\'e, CEA, CNRS, AIM, F-91191 Gif-sur-Yvette, France
    }
    
   \date{Received September xxx; accepted xxx}

 
  \abstract
   {The blue compact dwarf (BCD) galaxy I Zw 18 is one of the most metal-poor ($Z \sim 3\% \, Z_{\sun}$) star-forming galaxies known in the local Universe. Since its discovery, the evolutionary status of this system has been at the center of numerous debates within the astronomical community.}
   {We aim to probe and resolve the stellar populations of I Zw 18 in the near-IR using JWST/NIRCam's unprecedented imaging spatial resolution and sensitivity. Additionally, our goal is to derive the spatially resolved star formation history (SFH) of the galaxy within the last $1$ Gyr, and to provide constraints for older epochs.}
   {We used the point spread function fitting photometry package \texttt{DOLPHOT} to measure positions and fluxes of point sources in the $F115W$ and $F200W$ filters' images of I Zw 18, acquired as part of the JWST GTO ID 1233 (PI: Meixner). Furthermore, to derive I Zw 18's SFH, we applied a state-of-the-art color-magnitude diagram (CMD) fitting technique (\texttt{SFERA} 2.0), using two independent sets of stellar models: PARSEC-COLIBRI and MIST.}
   {Our analysis of I Zw 18's CMD reveal three main stellar populations: one younger than $\sim30$ Myr, mainly associated with the northwest star-forming (SF) region; an intermediate-age population ($\sim 100 - 800$ Myr), associated with the southeast SF region; and a red and faint population, linked to the underlying halo of the galaxy, older than $1$ Gyr and possibly as old as $13.8$ Gyr. The main body of the galaxy shows a very low star formation rate (SFR) of  $\sim 10^{-4} \; M_{\odot} \; {yr}^{-1}$ between $1$ and $13.8$ Gyr ago. In the last billion years, I Zw 18 shows an increasing trend, culminating in two strong bursts of SF around $\sim 10$ and $\sim100$ Myr ago. Notably, I Zw 18 Component C mimics the evolution of the main body, but with lower SFRs on average.}
   {Our results confirm that I Zw 18 is populated by stars of all ages, without any major gaps. Thus, I Zw 18 is not a truly young galaxy, but rather a system characterized by an old underlying stellar halo, in agreement with what has been found in other BCDs by similar studies. The low SF activity exhibited at epochs older than $1$ Gyr is in agreement with the ``slow cooking’’ dwarf scenario proposed in the literature, and could have contributed to its low metal content. The galaxy is now experiencing its strongest episode of star formation ($\sim 0.6 \, M_{\sun} \, yr^{-1}$) mainly located in the northwest region. A recent gravitational interaction between the main body and Component C is the most likely explanation for this starburst.}

   \keywords{Galaxy: evolution -- Galaxy: formation -- Galaxy: stellar content -- Galaxies: dwarf -- Galaxies: starburst -- Galaxies: star formation
               }

   \authorrunning{G. Bortolini et al.}
   \maketitle
%

\section{Introduction}
The blue compact dwarf (BCD) galaxy I Zw 18 is among the most enigmatic systems known in the local Universe, located at a distance of $\sim18.2$ Mpc \citep{Aloisi2007}. Since its discovery by \citet{Zwicky1966}, the nature of this system has sparked debates within the astronomical community. With its blue color \citep{Papaderos2002,Papaderos2012}, high gas content \citep{vanZee1998,Lelli2012}, and extreme metal deficiency ($Z \sim 1/30 \, Z_{\sun}$; \citealt{Searle1972,Lequeux1979,Skillman1993,Garnett1997,Vilchez1998,Izotov1999,Lecavelier2004}), I Zw 18 does indeed resemble a young galaxy at high redshift forming its first generation of stars \citep{Kunth1995,Izotov1999young,Izotov2004}. However, over the years, some authors have proposed an alternative interpretation. They suggested that I Zw 18 could be an old system that has formed stars continuously for a prolonged period of time, but at a rate too low to efficiently enrich its interstellar medium (ISM), the so-called slow cooking dwarf scenario \citep[see][]{Garnett1997,Legrand2000,Legrand2001}.   

\begin{figure}
  \centering
  \includegraphics[width=0.4\textwidth]{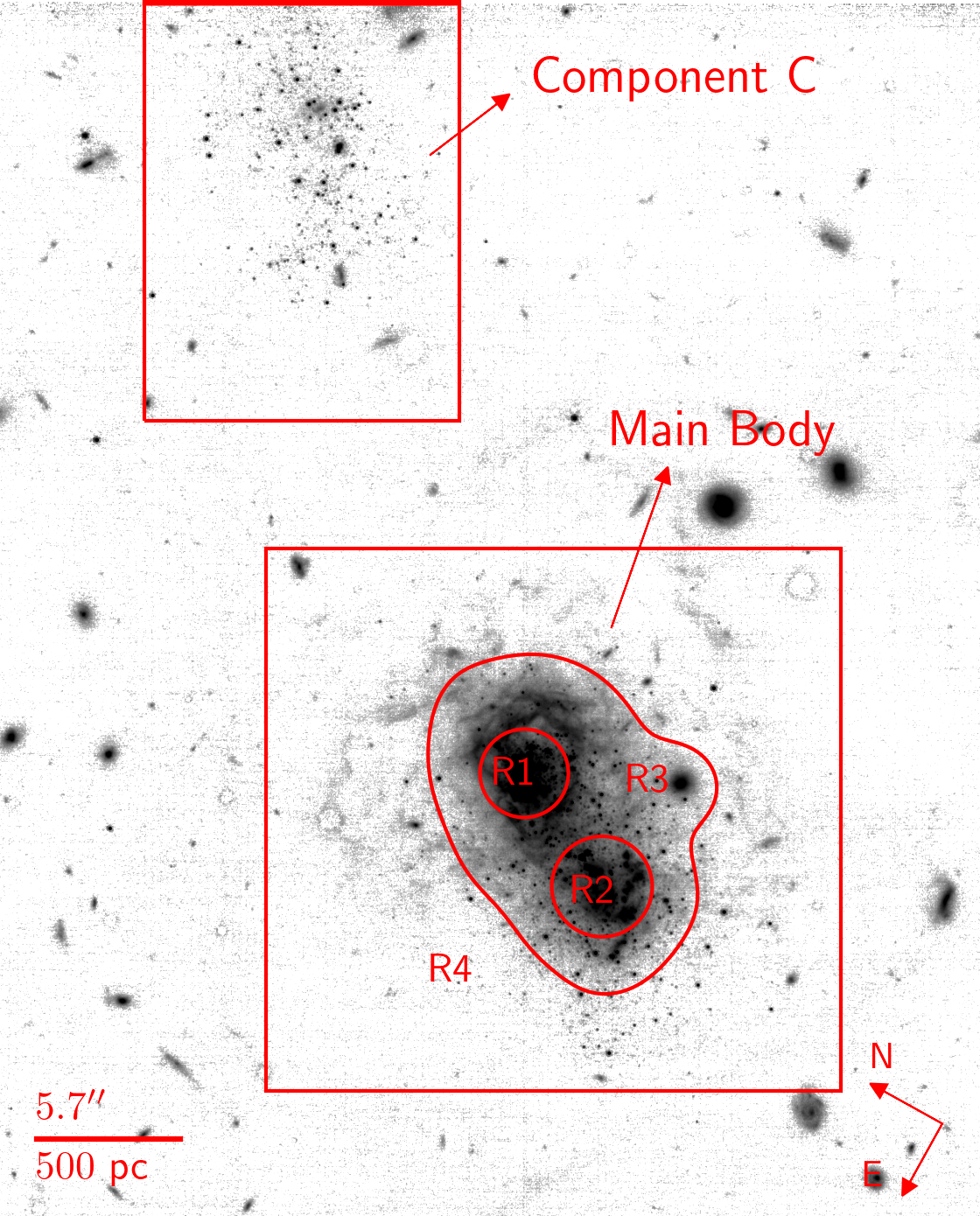}
  \caption{$F200W$ $(\sim K)$ JWST/NIRCam image cutout displaying I Zw 18's main body (southeast) and Component C (northwest). We divided the main body of the galaxy in four different regions, going from the two inner star-forming regions (labeled R1 and R2, respectively), a middle section (R3), and the outermost region (R4). The Component C is also labeled in the image. In the bottom-left corner is shown a physical scale corresponding to $\sim 5.7\arcsec$ ($\sim 500$ pc).}
  \label{F200W_cutout}
\end{figure}

I Zw 18 is comprised of two active star-forming (SF) regions: one located in the northwest (referred to as NW) and the other in the southeast (referred to as SE). These areas are surrounded by a bright nebular emission characterized by a complex and extensive filament structure \citep{Davidson1989,Dufour1990,Hunter1995,Dufour1996HST,Papaderos2002,Hirschauer2024}. Adding to the complexity of this system, I Zw 18 also has a faint companion (known in the literature as Component C) located at $\sim 22\arcsec$ from the main body (see Fig. \ref{F200W_cutout}). Many authors (e.g., \citealt{Dufour1990,Papaderos2002}, among others) describe I Zw 18 as a ``fragmented association of \ion{H}{i} clouds'', with Component C in the northwest and with an \ion{H}{i} tail extending to the southeast up to $1'$ \citep{vanZee1998,Lelli2012}. The physical association of Component C with I Zw 18 was first established by \citet{Dufour1996C} using \ion{H}{$\alpha$} and \ion{H}{$\beta$} emission lines. This association was later confirmed by 21 cm observations, which found that both I Zw 18 and Component C share a common extended \ion{H}{i} envelope \citep{vanZee1998,Lelli2012}. \citet{vanZee1998} and \citet{Lelli2012} also constrained the  dynamical mass of the system between $2-3 \times 10^{8} \, M_{\sun}$.

The most reliable method to estimate the star formation history (SFH) of nearby galaxies involves analyzing the color-magnitude diagram (CMD) of their resolved stellar populations. The advent of the Hubble Space Telescope (HST) launched in April 1990, together with the development of modern photometry reduction routines \citep{Stetson1987,Dolphin2000}, made it possible to resolve single stars and construct the CMD of both I Zw 18's main body and Component C. However, I Zw 18's distance and intrinsic compactness make stellar population studies of this galaxy rather challenging, with current photometry reaching only the brightest stellar evolutionary phases.   

The first I Zw 18 CMD was derived in the optical by \citet{Hunter1995} from deep HST/WFPC2 observations. They found a well-populated main sequence (MS) of massive stars with ages around 10 Myr, but no evidence of an older population. The same conclusion was reached a year later by \citet{Dufour1996HST} using independent HST/WFPC2 images. Later studies by \citet{Aloisi1999} and \citet{Ostlin2000} identified some candidate asymptotic giant branch (AGB) stars that constrain the age of the galaxy to $0.5-1$ Gyr. However, the authors assumed a distance of $\sim 10$ Mpc (estimated from its pure Hubble flow recessional velocity), which we now know to be heavily underestimated. This underestimation of the distance consequently led to an overestimation of the age of the galaxy.\par

A significant step forward was made possible with the Advance Camera for Surveys (ACS) mounted on board HST in the early 2000s. The much higher sensitivity of ACS allowed point source photometry to be extended to magnitudes as faint as $\sim 29$ mag in the I band, theoretically enabling the detection of faint and old red giant branch (RGB) stars. With ACS images, \citet{Izotov2004} confirmed the presence of a young MS of blue and red super-giant stars ($10 <$ Myr age $< 100$ Myr), but also the presence of an evolved population of AGB stars ($100 <$ Myr age $< 500$ Myr). Nevertheless, they found no evidence of RGB stars, supporting the idea that I Zw 18 is a bona fide young galaxy. However, many other authors exploited the capabilities of ACS, pushing point source photometry to fainter magnitudes and improving the CMD analysis. In particular, \cite{Momany2005}, \cite{Aloisi2007}, \cite{ContrerasRamos2011}, and \cite{Annibali2013}, suggested that I Zw 18 is older than $1$ Gyr. Their deep CMDs showed hints of an underlying population of red and faint RGB stars. Despite all the progresses made in the last 30 years, a final consensus on the age of I Zw 18 remains elusive. Today, almost 30 years after the HST launch, the James Webb Space Telescope (JWST) offers us the unique opportunity to study I Zw 18's stellar populations with unprecedented resolution and depth. Equipped with a $\sim 6.5$ meter diameter mirror and specialized sensitivity in the near-infrared (IR) spectrum, JWST stands as the perfect instrument to probe I Zw 18's old stellar population. As low- and intermediate-mass stars age and evolve past their MS phase, they do indeed start to emit more strongly in the redder and IR wavelengths.\par

Here we present our study of I Zw 18's main body and Component C $1.1$ $\mu$m  $-$ $2.0$ $\mu$m  vs. $2.0$ $\mu$m CMDs, together with the analysis of the spatial distributions of their main stellar populations. Furthermore, we apply a state-of-the-art CMD fitting technique \citep{Bortolini2024} using two independent sets of stellar models, PARSEC-COLIBRI \citep{Bressan2012,Marigo2017} and MIST \citep{Dotter2016}, to reconstruct the spatially resolved SFH of the galaxy within the look-back time reached by our data. Unveiling the SFH of I Zw 18 is a crucial step to understand the underlying mechanisms responsible for its extremely low metal abundance, as well as shedding light on star formation processes in low metallicity environments. \par

In this paper, we present the observations and data reduction in Sect. \ref{sec:Observations and Data Reduction}, then describe the stellar populations and their spatial distribution in Sect. \ref{sec:Stellar populations}.  We introduce the SFH analysis in Sect. \ref{sec:Star Formation History}, followed by a discussion of the results in Sect. \ref{sec:Discussion}. Finally we present the summary and conclusions in Sect. \ref{sec:Summary and Conclusions}.

\section{Observations and data reduction}\label{sec:Observations and Data Reduction}

\subsection{Data}
Both I Zw 18 main body and Component C have been imaged with NIRCam in four filters ($F115W$, $F200W$, $F356W$ and $F444W$) as part of the JWST GTO program (Program ID: 1233; PI: M. Meixner). For a detailed presentation and discussion of the observational strategies and data reduction, we refer the reader to \cite{Hirschauer2024}. Due to the observational strategy \citep{Hirschauer2024}, a small portion of Component C in the northwest falls outside the detector's field of view. The detector's boundary intersects an isophote with a surface brightness of approximately $24.5$ mag arcsec$^{-2}$. Therefore, all global properties of Component C recovered in the analysis (e.g., stellar mass) should be intended as lower limits. In the following, we utilize only the short-wavelength $F115W$ and $F200W$ images (see Fig. \ref{F200W_cutout}). These filters are particularly well-suited for conducting point spread function (PSF) photometry fitting thanks to their superior spatial resolution compared to longer wavelength filters. This enhanced resolution allows for the detailed observation of stellar populations within the galaxy, enabling an accurate analysis of its stellar constituents.


\subsection{Photometric reduction}
\label{sec:Photometric reduction}

To measure positions and fluxes of point-like sources on the whole field covered by the NIRCam observations, we performed PSF-fitting photometry in both $F115W$ ($\sim J$) and $F200W$ ($\sim K$) images using the NIRCam module of the \texttt{DOLPHOT} 2.0 package \citep{Dolphin2000,Dolphin2016,Weisz2024}\footnote{http://americano.dolphinsim.com/dolphot/}. We used the $F200W$ stage 3 drizzled \texttt{i2d.fits} image as astrometic reference to find point sources, and the single stage 2 frames processed with the JWST calibration pipeline \citep[see ][]{Hirschauer2024} to measure the flux of each identified point source. Before running the actual photometry, we masked out bad pixels using the built-in \texttt{DOLPHOT} \texttt{NIRCammask} routine. For photometric parameters, after extensive tests runs, we adopt the ones listed in Table \ref{table:Dolphot parameters}. The output photometry from \texttt{DOLPHOT} is on the calibrated Vega magnitude scale\footnote{https://jwst-docs.stsci.edu/jwst-near-infrared-camera/nircam-performance/nircam-absolute-flux-calibration-and-zeropoints}. Finally, we inspected and culled the \texttt{DOLPHOT} ``raw'' output catalog to remove possible interlopers (i.e., extended objects, blended sources, and spurious detections) while aiming to retain the largest number of bona fide stars. Thus, we required a signal-to-noise ratio $>2$, $|$\texttt{sharp}$|< 0.3$, \texttt{crowd}$< 0.4$ in each filter. The sharpness parameter quantifies the extent to which a source is either sharply peaked or broadly distributed in comparison to the PSF, and helps in the elimination of image artifacts and background galaxies. The crowding parameter gauges the increase in brightness a source would have experienced if nearby stars on the sky had been fitted independently rather than simultaneously. We also retain only sources with \texttt{objectype} $\leq 1$ \citep[see][for an in deph description of the output parameters]{Dolphin2000}. Due to Component C's proximity to the edge of the detector, we paid particular attention to excluding sources that were too close to the edge from our final photometric catalog to avoid possible spurious detections. The precise limit beyond which we could trust our photometry was determined through extensive artificial star tests (see Sect. \ref{sec:Artificial star tests} for more details). Moreover, to account for potential distant background galaxies which appear point like in the data, we analyzed the remaining part of the image (i.e., excluding the areas covered by I Zw 18's main body and Component C, see Fig. \ref{F200W_cutout}) using the same \texttt{DOLPHOT} parameters and photometric cuts described above. According to this test, we anticipate an average contamination of up to $1\%$ from background galaxies in our final catalog at $m_{F115W} - m_{F200W} \sim 1$ and $m_{F200W} \sim 27$, where the contamination of the RGB is expected to be most pronounced. Thus, we are confident that this possible contamination do not affect the main conclusions of the analysis. Our final catalog comprises  of approximately 900 and 300 bona fide stars for I Zw 18's main body and Component C respectively.
A similar analysis, but using a different PSF-photometry package (\texttt{STARBUGII}, \citealt{Nally2023}), has been performed by \cite{Hirschauer2024} on the same data, including also the $F356W$ and $F444W$ filters. A detailed comparison between the \texttt{DOLPHOT} and the \texttt{STARBUGII} photometry will be the subject of a forthcoming paper.

\begin{table}
\caption{\texttt{DOLPHOT} parameters.}              
\label{table:Dolphot parameters}      
\centering                                      
\begin{tabular}{c c c c c c c c}          
\hline\hline                        
  & Parameter & & & & & Value &  \\    
\hline                                   
      & \texttt{img\_shift} & & & & & 0 0 &   \\      
      & \texttt{img\_xform} & & & & &1 0 0 &   \\
      & \texttt{img\_RAper} & & & & & 2 &  \\
      & \texttt{img\_Rchi} & & & & & 1.5 &   \\
      & \texttt{img\_RSky0} & & & & &  15 &    \\
      & \texttt{img\_RSky1} & & & & &  35  &  \\
      & \texttt{img\_RSky2} & & & & &  3 10 &  \\
      & \texttt{img\_RPSF} & & & & & 15 &  \\
      & \texttt{img\_apsky} & & & & & 20 35 &  \\
      & \texttt{RCentroid} & & & & & 2 &  \\
      &\texttt{SigFind} & & & & & 3.0 &  \\
      &\texttt{SigFindMult} & & & & & 0.85 &  \\   
      &\texttt{SigFinal} & & & & & 3.5 &  \\         
      &\texttt{MaxIT} & & & & & 25  &  \\            
      &\texttt{PSFPhot} & & & & & 1  &  \\           
      &\texttt{PSFPhotIt} & & & & & 2    &  \\       
      &\texttt{FitSky} & & & & & 2   &  \\           
      &\texttt{SkipSky} & & & & & 1   &  \\         
      &\texttt{SkySig} & & & & & 2.25   &  \\        
      &\texttt{NoiseMult} & & & & & 0.10  &  \\      
      &\texttt{FSat} & & & & & 0.999  &  \\          
      &\texttt{PosStep} & & & & & 0.25  &  \\ 
      &\texttt{RCombine} & & & & & 1.5 &  \\          
      &\texttt{SigPSF} & & & & & 5.0  &  \\
      &\texttt{UseWCS} & & & & & 2 &  \\
      &\texttt{SecondPass} & & & & & 5 &  \\
      &\texttt{PSFres} & & & & & 1 &  \\
      &\texttt{ApCor} & & & & & 1 &  \\
      &\texttt{Force1} & & & & & 0 &  \\
      &\texttt{FlagMask} & & & & & 4 &  \\
      &\texttt{CombineChi} & & & & & 0 &  \\
      &\texttt{InterpPSFlib} & & & & & 1 &  \\
      &\texttt{NIRCamvega} & & & & & 1 &  \\
\hline                                            
\end{tabular}
\tablefoot{Parameters description at \\ http://americano.dolphinsim.com/dolphot/}
\end{table}

\subsection{Artificial star test}
\label{sec:Artificial star tests}

\begin{figure}
  \resizebox{\hsize}{!}{\includegraphics{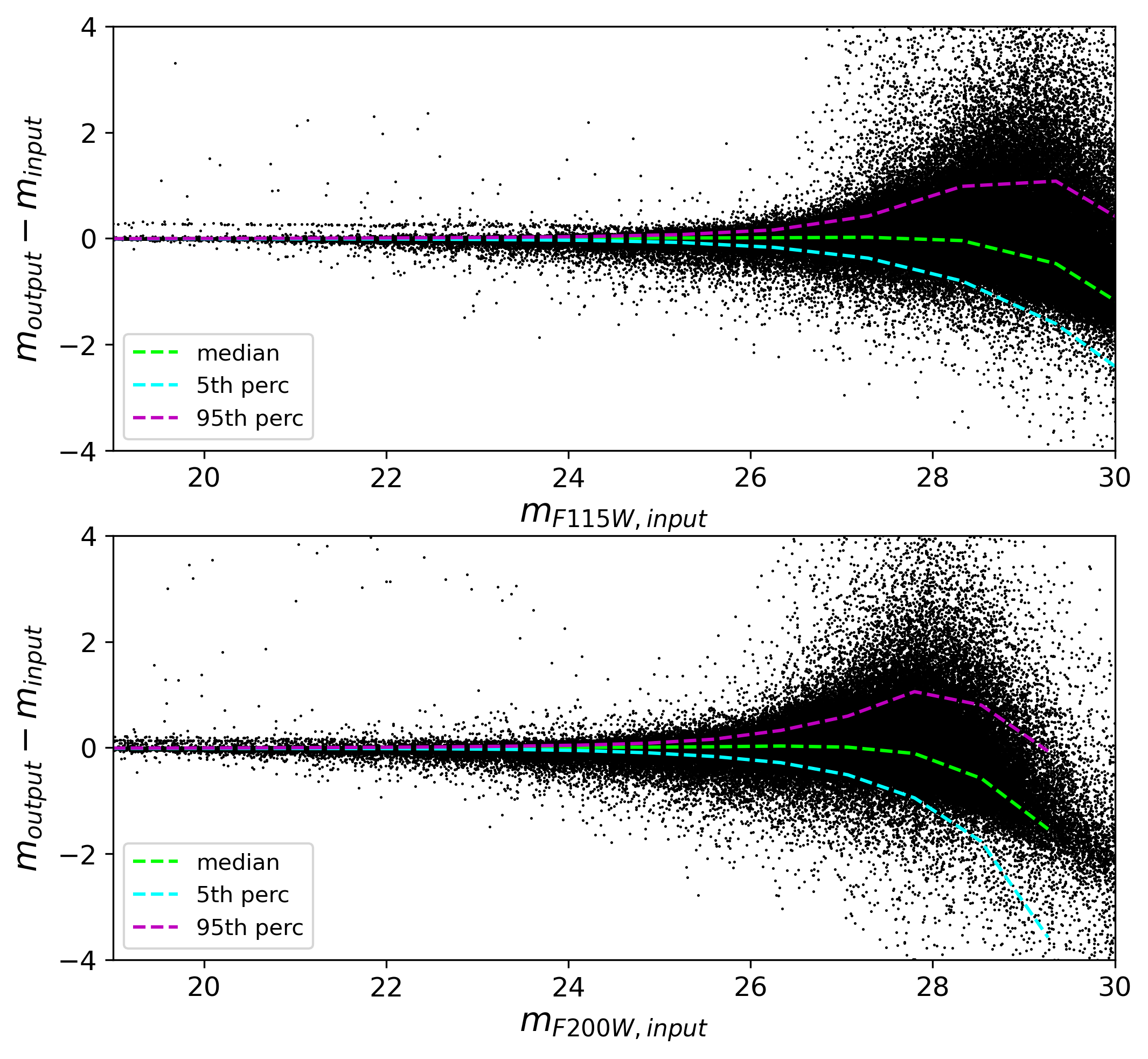}}
  \caption{$m_{output}-m_{input}$ vs. $m_{input}$ distribution of the artificial stars in $F115W$ (upper panel) and $F200W$ (bottom panel) filter. The green-dashed lines is the median of the distribution, while the cyan and the pink ones show the $5^{th}$ and $95^{th}$ percentile respectively.}
  \label{AST_magout_vs_magin}
\end{figure}

To estimate photometric errors and the completeness of our photometry we perform extensive artificial stars tests (ASTs). First, we generate a catalog of artificial stars, with input positions distributed accordingly to the galaxy's surface brightness and with input magnitude and color that span the range $m_{F200W} \in [18,30]$ and $m_{F115W} - m_{F200W} \in [-1,4]$. Each star is then injected into the actual image, which is then reprocessed with \texttt{DOLPHOT} to recover both real and artificial sources. It is important to stress that artificial stars are placed one at a time in the image, so as not to alter the intrinsic crowding of the target. An input artificial star is considered recovered only if its output magnitude is not $0.75$ mag brighter than the input value (i.e., recovered with less than twice the input flux). We then proceed to clean the artificial stars catalog applying the same photometric quality cuts applied to the data (see Sect. \ref{sec:Photometric reduction}). In galaxies with high density gradients, this approach allows us to get the best possible estimation of the completeness and the photometric errors as a function of magnitude, color, and position within the galaxy \citep{Cignoni2016,Sacchi2018}. In total we simulate $\sim 1.5$ million artificial stars. Fig. \ref{AST_magout_vs_magin} shows the $m_{output}-m_{input}$ vs. $m_{input}$ distribution of our artificial stars in both the $F115W$ and $F200W$ filters.

\begin{figure}
  \resizebox{\hsize}{!}{\includegraphics{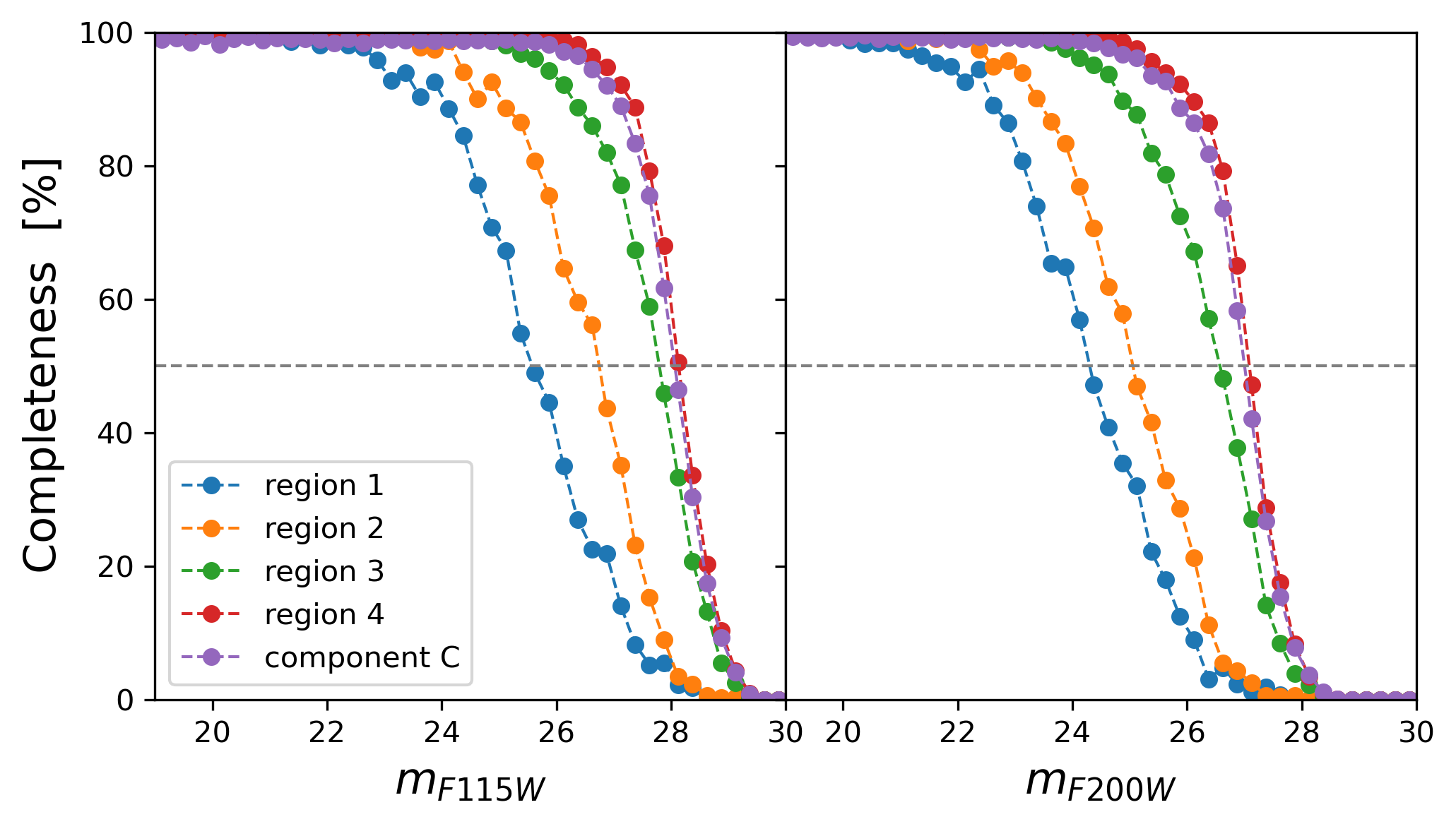}}
  \caption{Averaged photometric completeness from our AST, as a function of magnitude in the $F115W$ (left panel) and $F200W$ (right panel) filters. Different regions of the galaxy (see Fig. \ref{F200W_cutout}) are highlighted in different colors (see legend in the bottom-left corner). The dashed gray line marks the $50\%$ completeness limit.}
  \label{AST_regions_completeness}
\end{figure}

Given the high stellar density gradient and complex morphology of I Zw 18, we divided the field covered by our observations in five distinct spatial regions (see Fig. \ref{F200W_cutout}), and plot their averaged photometric completeness estimated from our ASTs in Fig. \ref{AST_regions_completeness}. Region 1 (R1) and region 2 (R2) are the innermost parts of the galaxy main body, which encompass the NW and SE SF regions respectively. Region 3 (R3) and region 4 (R4) are the middle and outermost sections of I Zw 18's main body. Finally, an additional region is defined that covers Component C. The plot shows how, in both filters, the averaged completeness goes down below $50\%$ at increasingly higher magnitudes (i.e., the photometry reaches fainter stars) going from more crowded regions (e.g., the ``center'' of the galaxy) to less crowded ones (e.g., the outskirts).    

\section{Stellar populations}\label{sec:Stellar populations}
In this Section, we analyze the near-IR CMDs of I Zw 18's main body and Component C and the spatial distribution of their stellar populations, probing different epochs in the lifetime of the galaxy.

\label{sec:color-magnitude diagram}
\subsection{Color-magnitude diagram}
\begin{figure*}
\centering
  \includegraphics[width=17cm]{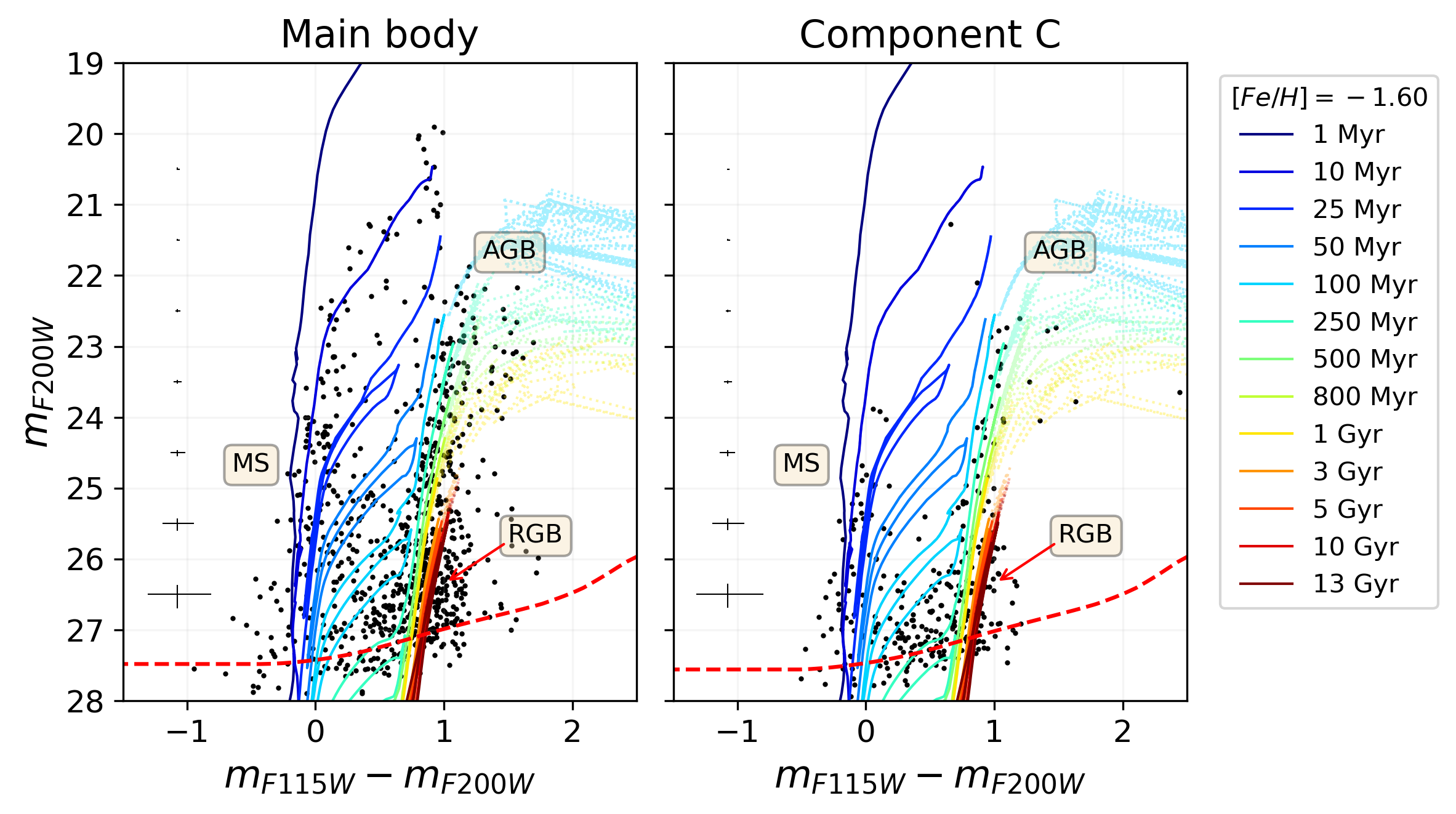}
  \caption{$m_{F200W}$ vs. $m_{F115W}-m_{F200W}$ color-magnitude diagrams of I Zw 18's main body (left panel) and Component C (right panel). Different stellar evolutionary phases are visible (see discussion in Sect. \ref{sec:color-magnitude diagram}). PARSEC-COLIBRI isochrones (assuming $[Fe/H] = -1.60$, $(m-M_{0}) = 31.3$ and $E(B-V) = 0.032$) are over-plotted and color-coded by their age (see the legend in the upper right corner). The main stellar evolutionary phases are labeled. Error bars on the left-hand side represent photometric uncertainties at various magnitudes, assuming $m_{F115W}-m_{F200W} = 1$. The red-dashed line indicates the $50\%$ completeness limit.}
  \label{I Zw 18_CMD}
\end{figure*}

Fig. \ref{I Zw 18_CMD} shows the $m_{F115W}-m_{F200W}$ vs. $m_{F200W}$ CMDs of I Zw 18's main body (left panel) and Component C (right panel), after applying the photometry quality cuts discussed in Sect. \ref{sec:Photometric reduction}. Photometric errors at different magnitude bins assuming $m_{F115W}-m_{F200W} = 1$  (error bars on the left-hand side) and the $50\%$ completeness limit (red-dashed line) estimated from our AST (see Sect. \ref{sec:Artificial star tests}) are also shown. To assist the reader's eye in exploring the various stellar populations present in the diagrams and their respective ages, we plot a set of PARSEC-COLIBRI isochrones\footnote{http://stev.oapd.inaf.it/cgi-bin/cmd} \citep{Bressan2012,Tang2014,Chen2014,Chen2015,Marigo2017,Pastorelli2020} on top of the CMDs, assuming a metallicity of $[Fe/H] = -1.60$, a distance modulus of $(m-M_{0}) = 31.3$ \citep{Aloisi2007}, and a foreground reddening of $E(B-V) = 0.032$ \citep{Schlafly2011}. The different isochrones are color-coded according to their age, going from younger populations (bluer colors) to older populations (reddish colors). The CMDs shows a wide range of stellar populations without any noticeable gaps, suggesting a star formation which has spanned continuously from ancient to recent epochs (see also Sect. \ref{sec:Star Formation History}). We can clearly identify two well separated bright features in the main body's diagram: a blue ``plume'' at $m_{F115W}-m_{F200W} \sim 0.0$ populated by upper MS stars and core He-burning phase of intermediate and high-mass stars; a red ``plume'' at $m_{F115W}-m_{F200W} \sim 0.6-0.7$, populated by more evolved AGB stars. Few thermally pulsating AGB (TP-AGB) stars are also present, outlining a secondary red ``plume'' around $m_{F200W} \sim 23$ and $m_{F115W}-m_{F200W} \sim 1.5$. Below the bright red ``plume'', there is a fainter over-density of stars falling on top of isochrones older than $1$ Gyr. The majority of these stars are in their RGB phase, with the tip at approximately $m_{F200W} \sim 26$. The unparalleled sensitivity of JWST in the near-IR enables us to detect this relatively faint phase, going from its tip down to one magnitude deeper, just before reaching the completeness limit of our photometry. Thus, our data confirm that I Zw 18 is indeed an old system and not a young galaxy currently in the process of forming its first generation of stars \citep{Hunter1995,Izotov1999young,Izotov2004}. I Zw 18's companion (i.e., Component C) exhibits similar features to those described above, albeit generally less populated. 

\subsection{Spatial distribution}
\begin{figure*}
\centering
   \includegraphics[width=18cm]{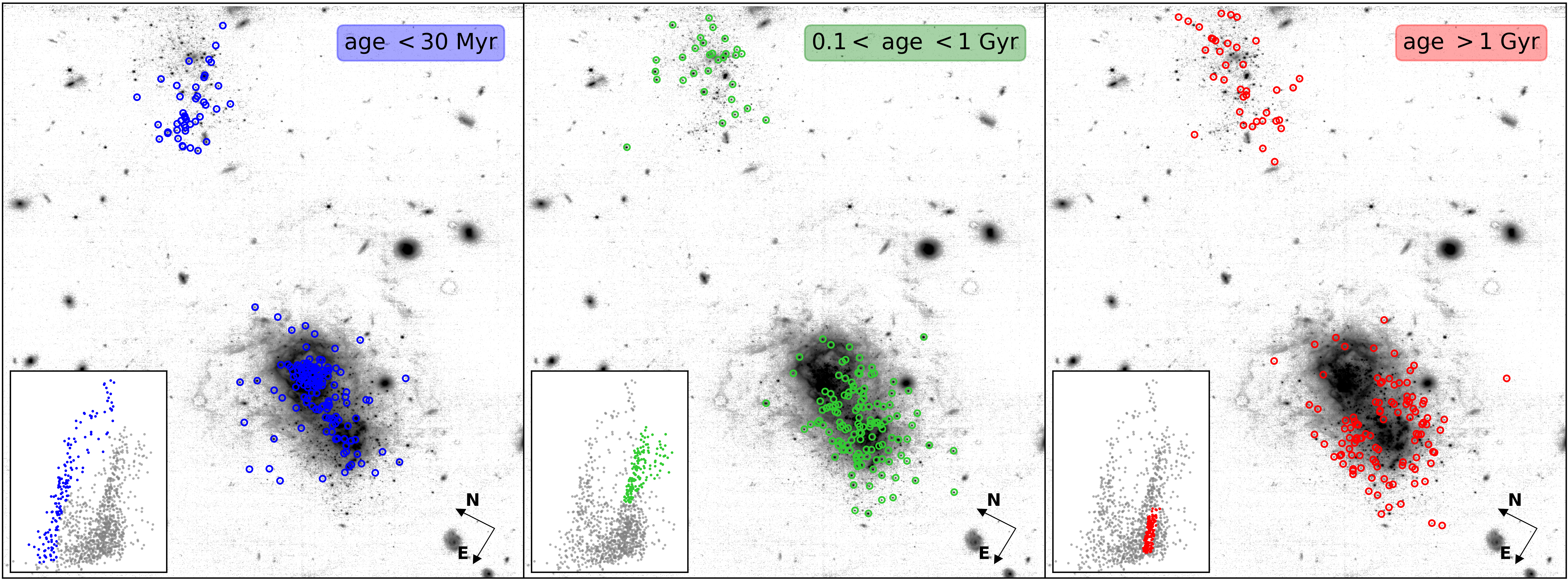}
     \caption{F200W image of I Zw 18 with the positions of three bona fide stellar populations being overplotted and selected in the CMD using PARSEC-COLIBRI isochrones. A population of upper MS stars younger than $30$ Myr (blue circles, left panel), AGB stars of intermediate ages between $0.1$ and $1$ Gyr (green circles, middle panel), and a population of RGB and low-mass AGB stars older than $1$ Gyr (red circles, right panel). The insert panels on the bottom-left show the respected position in the CMD of the selected bona fide stellar populations.}
     \label{fig:spatial distribution}
\end{figure*}

To gauge how different stellar populations, tracing different epochs, are spatially distributed across the galaxy's main body and Component C, we selected three bona fide stellar populations with different age intervals, using PARSEC-COLIBRI isochrones as a guide (see bottom-left panels of Fig. \ref{fig:spatial distribution}). These populations include upper MS stars younger than $30$ Myr (blue circles), AGB stars with intermediate ages ranging from $0.1$ to $1$ Gyr (green circles), and RGB and low-mass AGB stars older than $1$ Gyr, and potentially as old as the Universe (red circles). Fig. \ref{fig:spatial distribution} illustrates their positions within the galaxy. In general young stars show a concentrated and clumpy distribution, while the intermediate and old stars display a more open distribution, as expected since stars migrate from their birth locations as they age due to orbital mixing. We see a great concentration of very young stars within the main body's NW SF region, while the SE region seems more populated by intermediate-age stars. This possibly suggests that SF has started in the SE part of the main body and then moved ``up'' toward the NW region. The older RGB and AGB stars are distributed along the outskirts of the main body, with a higher number found around the SE SF region. This finding seems to confirm the idea that the SE part of the main body is substantially older than the NW region, which is now going through its main burst of SF. Nevertheless, due to the intrinsic faintness at the distance to I Zw 18, these stars are very difficult to detect. Thus, we cannot rule out the possibility that the non-detection around the NW SF region and in the inner part of the main body is due to incompleteness of our photometry. Notably, Component C exhibits a mirrored spatial distribution of stars. In particular, its young stellar population is clustered toward the main body. This clustering, together with the common HI envelope \citep{vanZee1998,Lelli2012}, might indicate that a recent gravitational interaction between the main body and Component C is responsible for the current strong starburst experienced by the galaxy \citep{Kim2017}.  

\section{Star formation history}\label{sec:Star Formation History}
\subsection{CMD fitting method}
In this Section, we present I Zw 18's SFH derived through the synthetic CMD fitting method, originally pioneered by \citet{Tosi1991} and successfully exploited in the last 30 years to derive the SFH of galaxies in the Local Universe \citep{Gallart1999,Annibali2003,Dolphin2005,McQuinn2010,Weisz2014,Sacchi2018,Cignoni2019,Bortolini2024}. In particular, we use the CMD fitting routine \texttt{SFERA} 2.0  (Star Formation Evolution Recovery Algorithm, \citealt{Bortolini2024}). We provide here only a short description of \texttt{SFERA} 2.0. We refer the interested reader to \citet{Bortolini2024} for an in depth description of the routine.

The synthetic CMD method involves the comparison of the observed galaxy CMD, with 
a set of theoretical CMDs generated from some assumed stellar evolution models (i.e., isochrones). In particular we chose two independent and widely used sets of stellar isochrones: PARSEC-COLIBRI \citep{Bressan2012,Tang2014,Chen2014,Chen2015,Marigo2017,Pastorelli2020} and MIST \citep{Dotter2016,Choi2016}\footnote{https://waps.cfa.harvard.edu/MIST/index.html}.  
First, we generate a library of ``simple'' synthetic stellar populations with all possible combinations of ages ranging from 1 Myr to 14 Gyr in steps of $\Delta\log(t) = 0.25$, and metallicities within $[Fe/H] \in [-2.00,-1.50]$ in steps of 0.1 dex. We note that we adopted the approximation $[Fe/H] = \log(Z/Z_{\sun})$, with $Z_{\sun} = 0.0152$ for PARSEC-COLIBRI \citep{Caffau2011} and $Z_{\sun} = 0.0142$ for MIST \citep{Asplund2009}. The choice of equal logarithmic age bins is to account for the increasingly lower time resolution at older epochs. The metallicity interval goes from the lowest value made available by both PARSEC and MIST models to the current value inferred from spectroscopic observations of the galaxy \ion{H}{ii} regions \citep{Skillman1993,Izotov1999}.
We also assumed a \citet{Kroupa2001} initial mass function between $0.2$ and $300$ $M_{\sun}$ and a $30\%$ of unresolved binaries.
We then construct the CMD associated with each ``simple'' synthetic stellar population (hereafter, basis synthetic CMDs), applying bolometric corrections based on stellar atmosphere and synthetic spectra models \citep{Conroy2010,Chen2019}. To mimic the observation conditions we then convolve the synthetic CMDs basis functions with the following data properties: photometric errors, blending and incompleteness derived from our AST (see Sect. \ref{sec:Artificial star tests}), distance modulus of $(m-M_{0}) = 31.3$ (derived by \citealp{Aloisi2007} with Cepheid variables), foreground reddening of $E(B-V) = 0.032$ \citep{Schlafly2011}. All the basic synthetic CMDs can now be linearly combined with appropriate weights to reproduce the observed CMD and recover its associated SFH. To identify the combination of weights that best reproduce the data, we bin both synthetic and observed CMDs, and compare the star counts in each color-magnitude grid cell following a Poissonian likelihood function \citep{Dolphin2002}. The minimization of the likelihood is performed by means of a genetic algorithm \citep[\texttt{Pikaia}, see][]{Charbonneau1995}. Finally, statistical uncertainties are computed with a bootstrap technique, that is, applying small random shifts in color and magnitude to the stars in the observed CMD and deriving again the SFH using \texttt{SFERA} 2.0. We then averaged all the solutions and take the 16, 50 and 84 percentiles of the distribution as the ``best-fit'' confidence interval. Systematic uncertainties are accounted for by deriving the SFHs using different color and magnitude binning in the minimization. Furthermore, the whole process is performed independently using synthetic CMDs generated from PARSEC-COLIBRI and MIST stellar models. For fixed age and metallicity values, the differences between the models predictions reflect different underlying stellar physics assumptions, giving us a more robust estimation of the possible systematic uncertainties in the fit. \par 

\begin{figure*}
\centering
   \includegraphics[width=17cm]{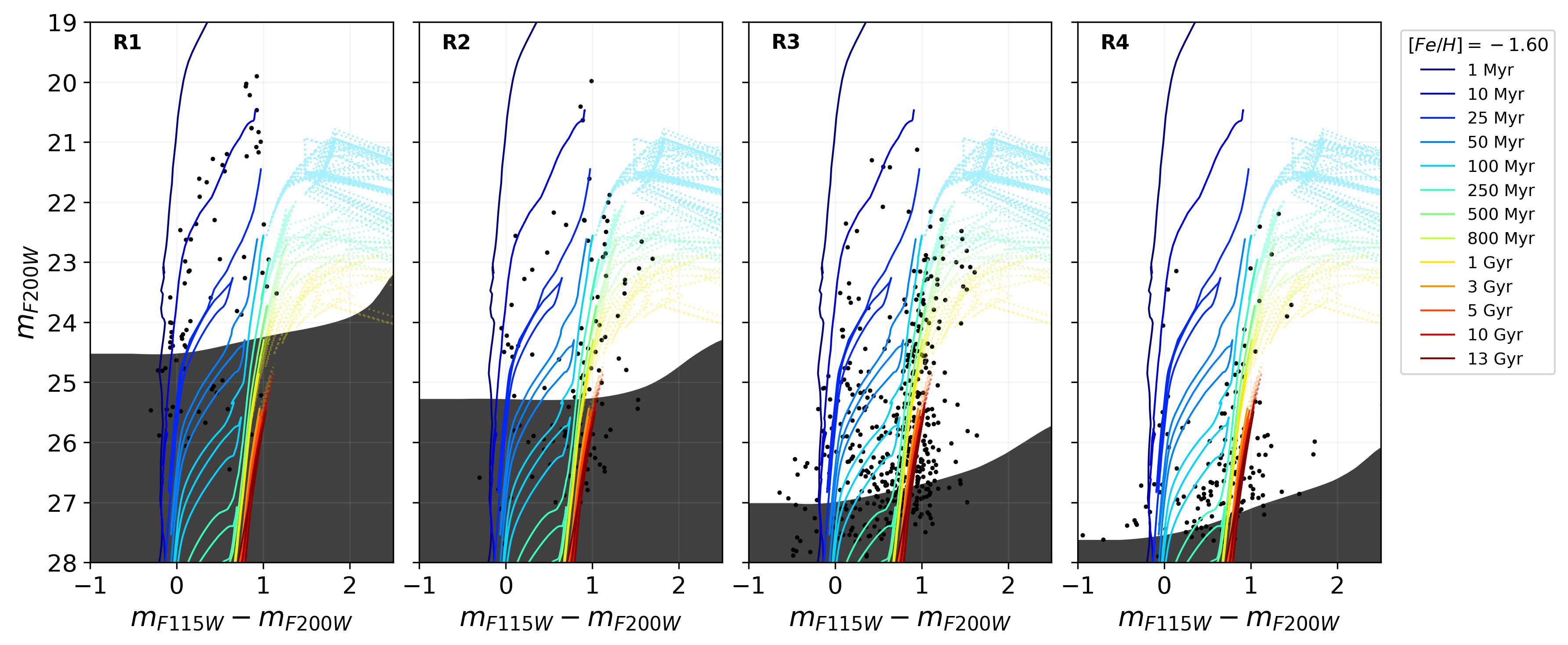}
     \caption{CMDs of the four different galaxy regions identified in the main body of I Zw 18 (see Fig. \ref{F200W_cutout}) and for which we independently recovered the SFH using \texttt{SFERA} 2.0. Region 1 (R1) and region 2 (R2) are the two innermost SF regions, in the northwest and southeastern part of I Zw 18's main body respectively. Region 3 (R3) and region 4 (R4) are the middle and outermost selections of the main body. Over-plotted on the CMDs and color-coded according to their age (see legend on the top-right corner) are some PARSEC-COLIBRI isochrones with $[Fe/H] = -1.60$. The darker shading obscures the region of the CMD that falls below our established $50\%$ completeness level.}
     \label{fig:regions_CMDS}
\end{figure*}

Before delving deep into the detailed discussion of I Zw 18's SFH, in Fig. \ref{fig:regions_CMDS} we show the CMDs of the four spatial regions in which we divided I Zw 18's main body (see Fig. \ref{F200W_cutout}) and for which we independently recover the SFH using \texttt{SFERA} 2.0. The CMDs are overlaid with a set of PARSEC-COLIBRI model isochrones, with $[Fe/H] = -1.60$ and ages as labeled in the legend between $10$ Myr and $10$ Gyr. The darker shading obscures the region of the CMD that falls below our fiducial $50\%$ completeness level. As already discussed in Sect. \ref{sec:Artificial star tests}, the photometry gets progressively ``deeper'' moving from region 1 and 2 (innermost part of the galaxy) to region 4, where we confidently detect stars as faint as $m_{F200W} \sim 27.5$. Already at a quick glance, the CMDs reveal profound differences. Both region 1 and 2 show CMDs dominated by very young ($10 - 500$ Myr) upper-MS and core He-burning stars. Unfortunately, the severe intrinsic crowding within these star-forming regions complicates the detection of faint, old stars, thereby hindering a comprehensive analysis of all stellar populations within these areas. Region 3 shows the most populated CMD among all the selected regions, with stellar populations of all ages. In particular, it displays a clear MS (even if less populated in the brightest part compared to region 1 and 2), a clear red ''plume'' around $m_{F115W}-m_{F200W} \sim 1$ populated by intermediate age core He-burning and AGB stars, but also a population of old RGB stars below $m_{F200W} \sim 26$. Region 4 still shows traces of blue MS and red AGB stars, along with a sparsely populated but still present RGB population, detected well above the $50\%$ completeness limit. Lastly, we also recover Component C's SFH from its CMD shown in the right panel of Fig. \ref{I Zw 18_CMD} (see Sect. \ref{sec:color-magnitude diagram}).

\subsection{Region 1}
\begin{figure*}
\centering
     \includegraphics[width=17cm]{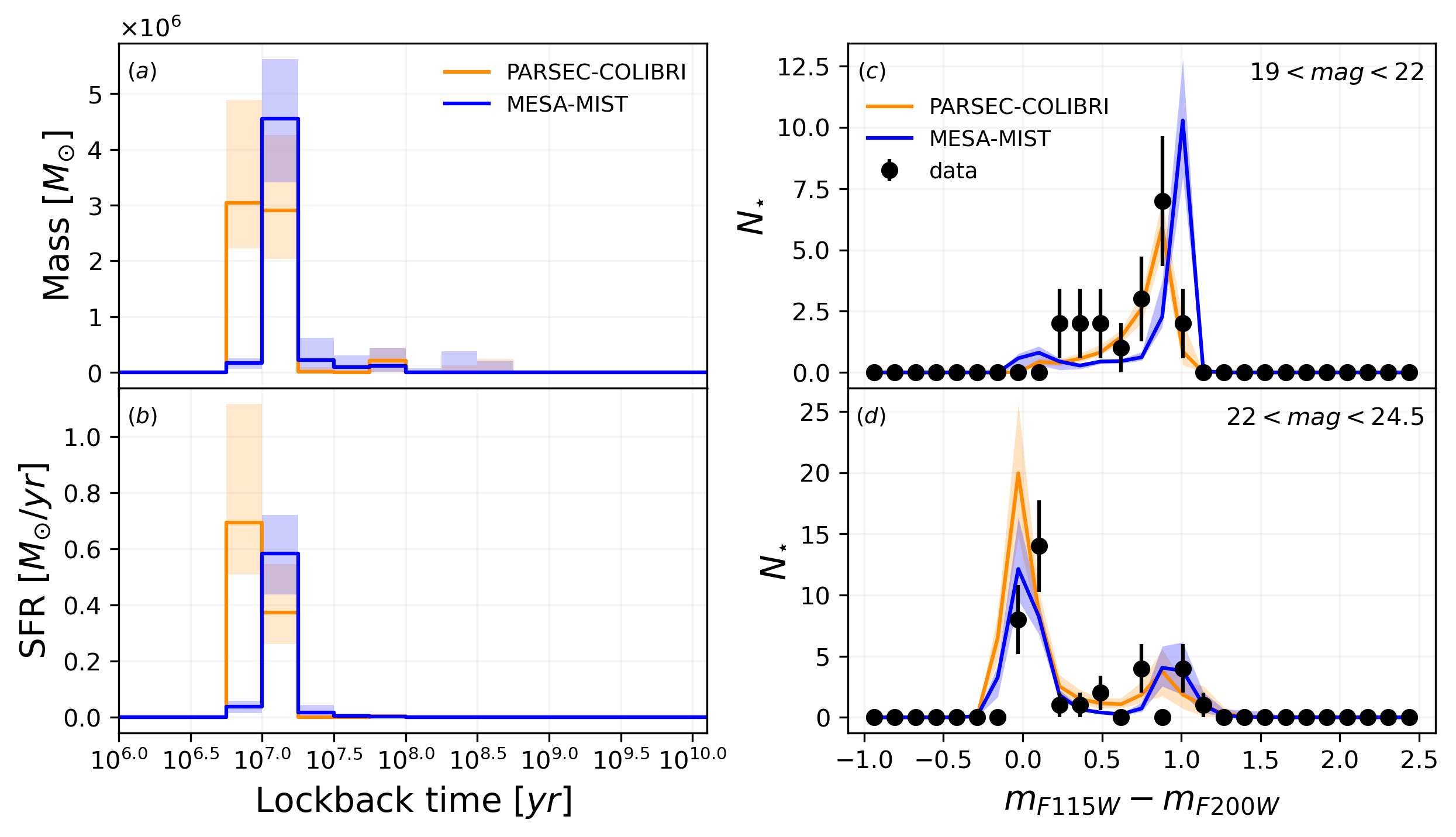}
     \includegraphics[width=17cm]{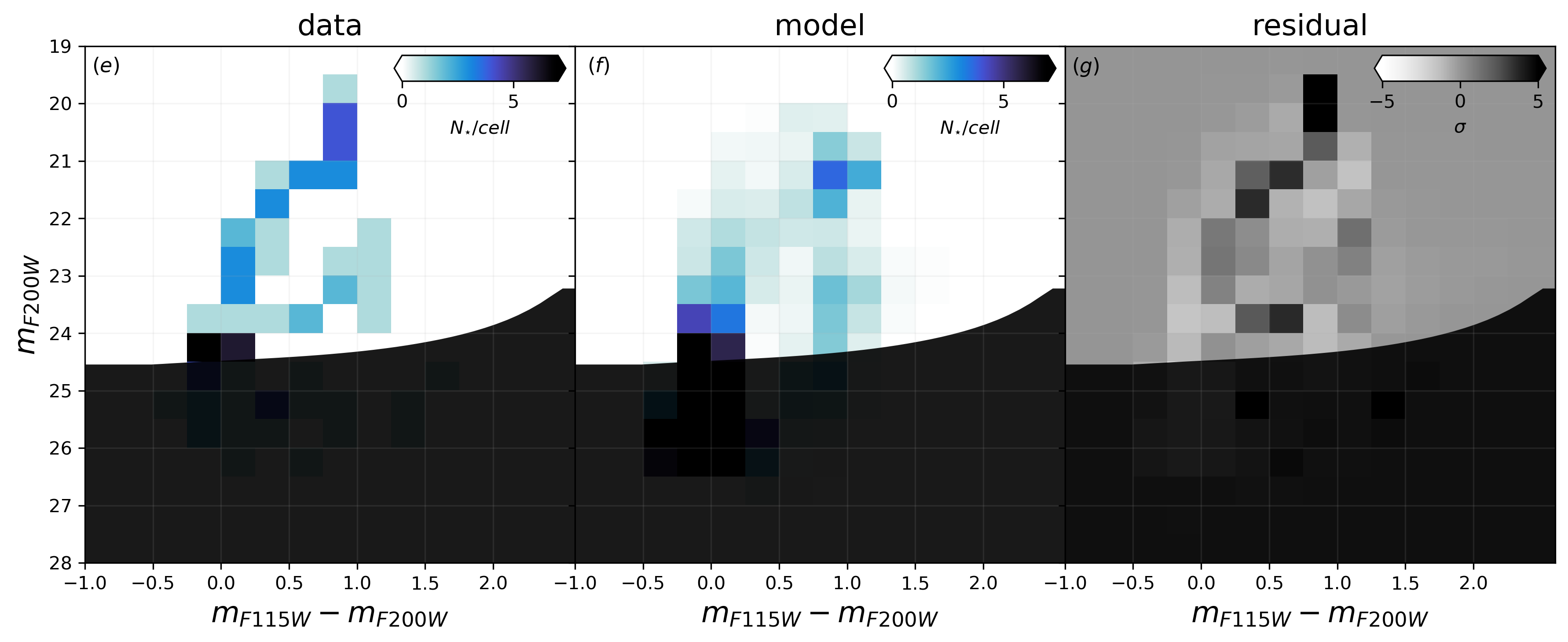}
     \caption{\texttt{SFERA} 2.0 results for region 1. Panels (a)-(b): recovered SFH and mass using PARSEC-COLIBRI (orange line) and MIST (blue line) isochrones. The shaded areas mark the 16-84 percentiles. Panels (c)-(d): Data (black points) vs. model CMD luminosity functions for different luminosity intervals (indicated on the upper-right). Again, the PARSEC-COLIBRI solution is in orange, while the MIST in blue. Panels (e)-(g): Observational CMD, \texttt{SFERA} 2.0 best fit model CMD and normalized residuals for each CMD cell ($data - model/\sqrt{model}$). The diagrams are color-coded according to the number of stars in each cell (see color-bar in the upper-right corner). The dark-shade masks the region of the CMD that is below our fiducial $50\%$ completeness limit. This region is excluded from the fitting procedure.}
     \label{fig:region1_results}
\end{figure*}

Region 1 encompasses the NW star-forming region and its corresponding \ion{H}{II} region. Due to the severe crowding conditions and the presence of very bright massive stars, our photometry reaches its $50\%$ completeness limit around $m_{F200W} = 24.5$. As can be seen from Fig. \ref{fig:regions_CMDS}, only young stars lie above this limit in the CMD, while stellar evolutionary phases that trace older epochs are all much fainter and all stay below our fiducial limit. Thus, in this region our look-back time is strictly limited to $\sim 500$ Myr. Fig. \ref{fig:region1_results} shows the results of our analysis for region 1. Panels (a) and (b) show the stellar mass formed and the SFR at different epochs (i.e., SFH), respectively. The results for the two independent sets of stellar models used to construct the synthetic CMD are displayed in orange (PARSEC-COLIBRI) and blue (MIST). The solid line is the median, while the shaded areas mark the 16 and 84 percentile of the different bootstraps solutions. As expected we recover a very strong burst of star formation (i.e., SFR $\sim 0.6 \, M_{\sun} \, yr^{-1}$) around $8-30$ Myr ago in both PARSEC and MIST best solutions. This burst of SF is quite strong compared to the current SFR exhibited by other Local Volume dwarf galaxies \citep{Tolstoy2009,McQuinn2010,Cignoni2019}, and even compared to other BCDs \citep{Crone2002,Garcia-Benito2012,Sacchi2016,Sacchi2021}. The high inferred SFR is necessary to account for the large number of blue super-giants present in the CMD. We would like to stress that, due to the limit imposed by the completeness of our photometry, we cannot put any constraints on the SFH of this region at epochs older than $\sim 500$ Myr, and that therefore this might not be its first episode of SF. Panel (c) and (d) show the color distribution of the data (black points), compared to the best model for the two sets of stellar models used, in two different magnitude intervals (see upper-right corner in the Fig.). Finally, Panels (e)-(g) show the observed CMD of region 1, the \texttt{SFERA} 2.0 best-fit model, and the residuals between the two. The CMDs are all binned and color coded according to the density of stars (see color-bar on the upper-right corner). The black-shaded area shows our photometry $50\%$ completeness limit. All the stars that fall below this limit have been excluded from the fitting procedure.    


\subsection{Region 2}
\begin{figure*}
\centering
     \includegraphics[width=17cm]{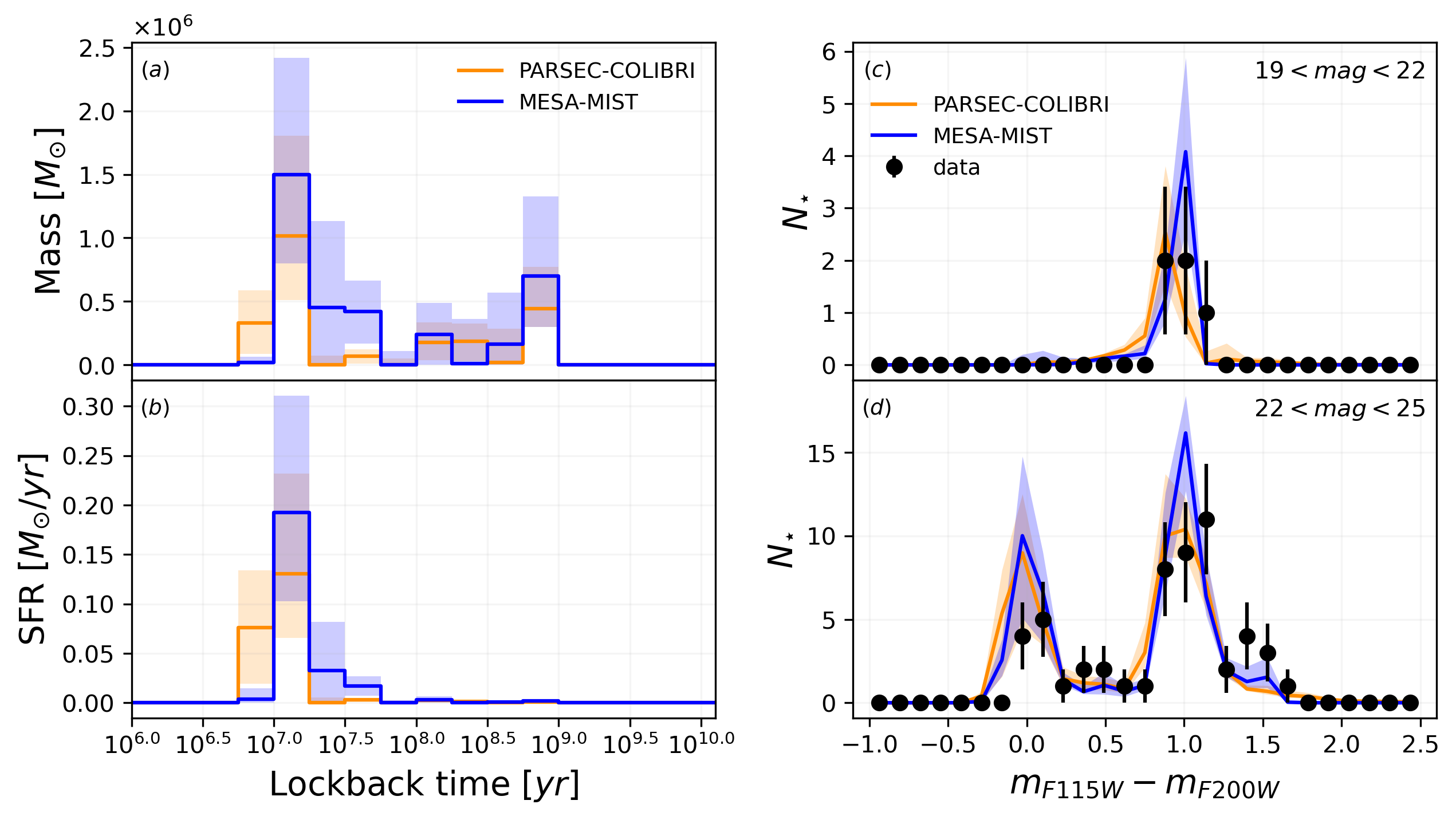}
     \includegraphics[width=17cm]{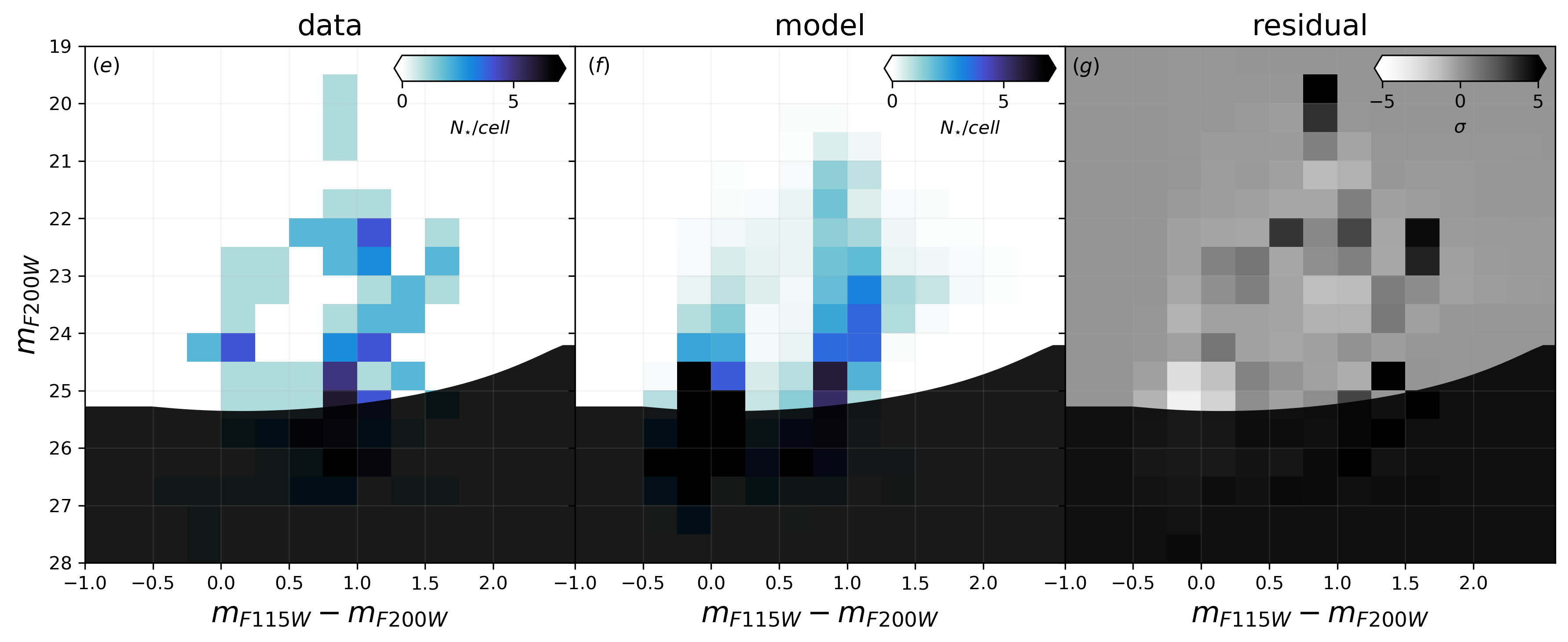}
     \caption{\texttt{SFERA} 2.0 results for region 2 (see Fig. \ref{fig:region1_results} for a detailed description of each panel).}
     \label{fig:region2_results}
\end{figure*}

Region 2 has been selected to study the SE star-forming region of the main body. As for region 1, the extreme crowding conditions of this region do not allow us to detect with confidence faint RGB stars that trace ancient epochs. Nevertheless, as can be seen in Fig. \ref{fig:regions_CMDS}, in this region we reach the $50\%$ completeness limit around $m_{F200W} = 25$. This enables us to probe this region's SFH up to $\sim 1$ Gyr ago. The results of our analysis are shown in Fig. \ref{fig:region2_results}. As region 1, also region 2 shows a dominant peak of  recent star-formation activity around $8-30$ Myr ago. However, region 2 also shows signs of activity at older epochs, having formed stars at a low but continuous rate since $1$ Gyr ago. Again, the depth of our photometry does not allow us to put constraints on the star-formation activity of this region at older epochs. As in region 1, both PARSEC-COLIBRI and MIST solutions agree within the uncertainties. The quality of the fit can be appreciated from the color distributions (Panels (c) and (d) of Fig. \ref{fig:region2_results}). In particular, we notice how our synthetic CMD is able to well-reproduce both the blue and red ``plumes'' present in the observed one. Again, Panels (e)-(g) show the observed, best model and residuals CMDs, binned accordingly to the density of points in each cell. 

\subsection{Region 3}
\begin{figure*}
\centering
     \includegraphics[width=17cm]{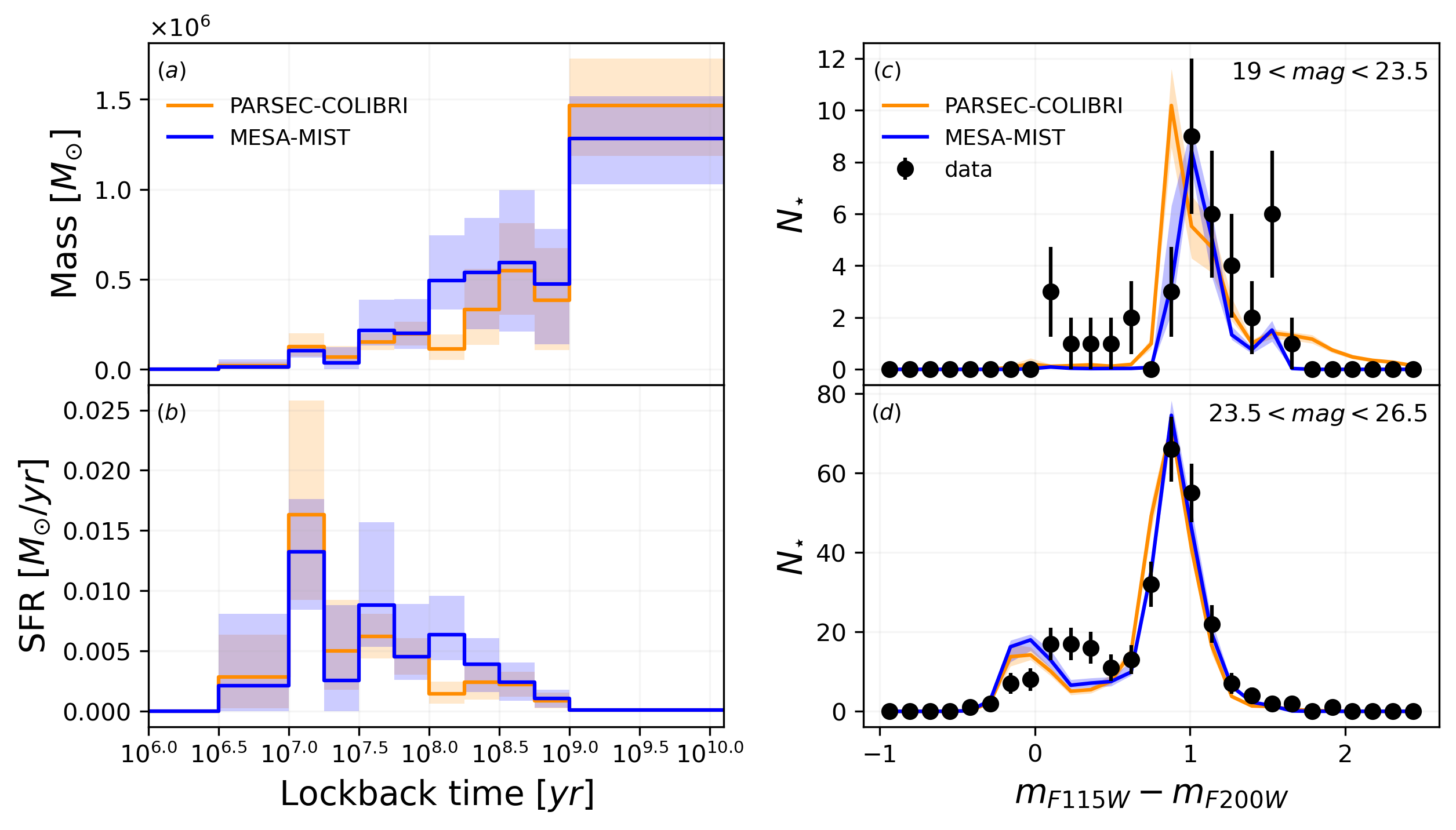}
     \includegraphics[width=17cm]{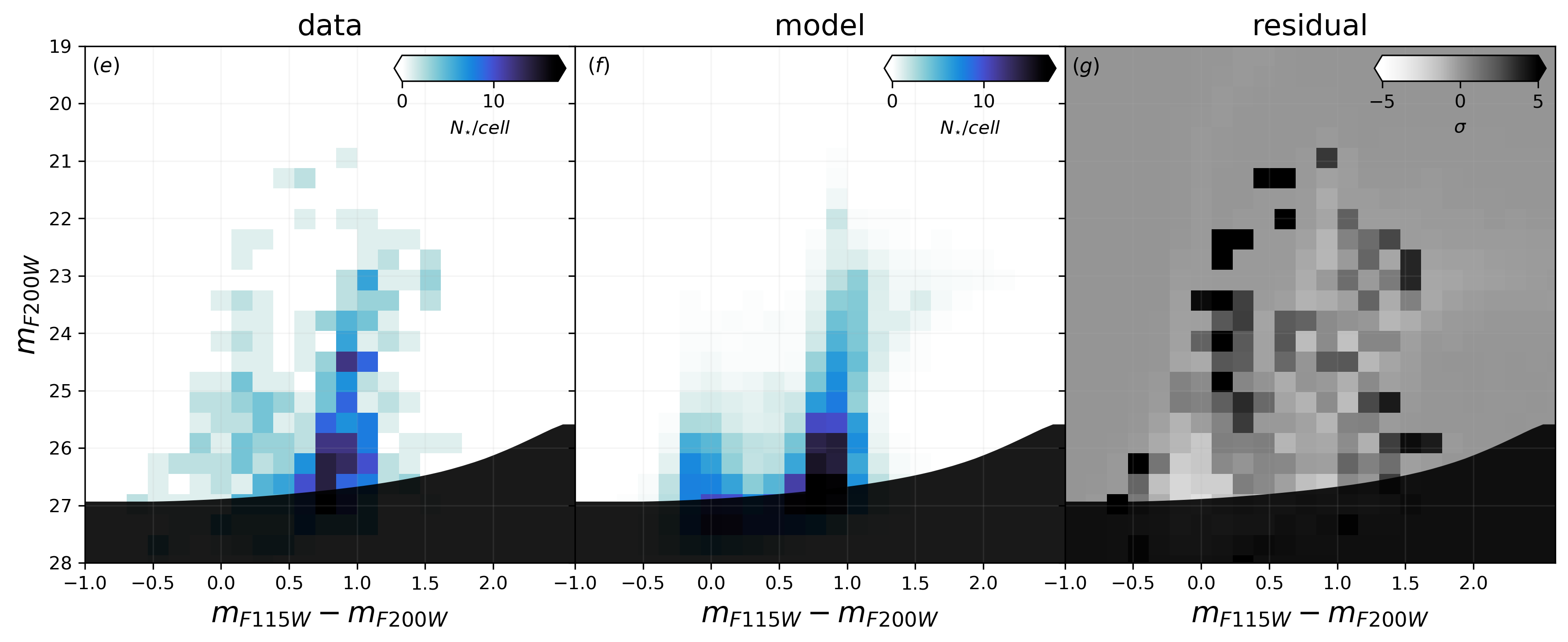}
     \caption{Same as Fig. \ref{fig:region1_results}, but for region 3 of the main body of I Zw 18.}
     \label{fig:region3_results}
\end{figure*}
As shown in Fig. \ref{fig:regions_CMDS}, region 3 displays a variety of stellar populations, tracing very different epochs in the life of the galaxy. Remarkably, we detect a clear population of RGB stars at $m_{F115W}-m_{F200W} \sim 0$ and $m_{F200W} \sim 26$. These stars are not a very good age indicator, due to the fact that they suffer from the age-metallicity degeneracy \citep{Gallart2005,Cignoni2010,Freedman2020}. To precisely trace the SF activity at older epochs one would need to resolve fainter stellar evolutionary phases, like the red clump (tracing 1-4 Gyr old stars) and the oldest main sequence turn-off (13.8 Gyr old stars, \citealt{Annibali2022}). Unfortunately, with the current generations of space telescopes this is not possible for galaxies beyond $\sim 4$ Mpc \citep{Annibali2022}. For this reason, in the fitting routine we adopted a single age bin from 1 to 14 Gyr. Nevertheless, we stress that the presence of these stars unequivocally implies that this region started forming stars at least 1 Gyr ago, and potentially even $13.8$ Gyr ago. Fig. \ref{fig:region3_results} shows our results for region 3 in detail. Both models agree that region 3 shows a very low star forming activity from 13.7 to 1 Gyr ago (SFR $\sim 10^{-4} \, M_{\sun} \, yr^{-1}$). After that, it seems to go through a steady increase in SF activity in the last billion years, reaching the peak activity at very recent times ($\sim 10$ Myr ago). Nonetheless, the bulk of the stellar mass of the region is locked in the oldest population of stars, for which we recover a lower limit for the stellar mass of $\sim 1.5 \times 10^{6} M_{\sun}$. The goodness of the fit can be appreciated from the color functions (Panels (c) and (d)) and from the comparisons of the observed and best-model CMDs (Panels (e)-(g)). Our model is able to reproduce remarkably well all the main features of the observed CMD at once (i.e., MS, AGB, RGB), with the exception of a few massive stars at the very bright end of the upper MS. These stars are unstable against strong stellar winds, fast rotation, and are characterized by a very short lifespan; thus, making them very challenging to model.

\subsection{Region 4}
\begin{figure*}
\centering
     \includegraphics[width=17cm]{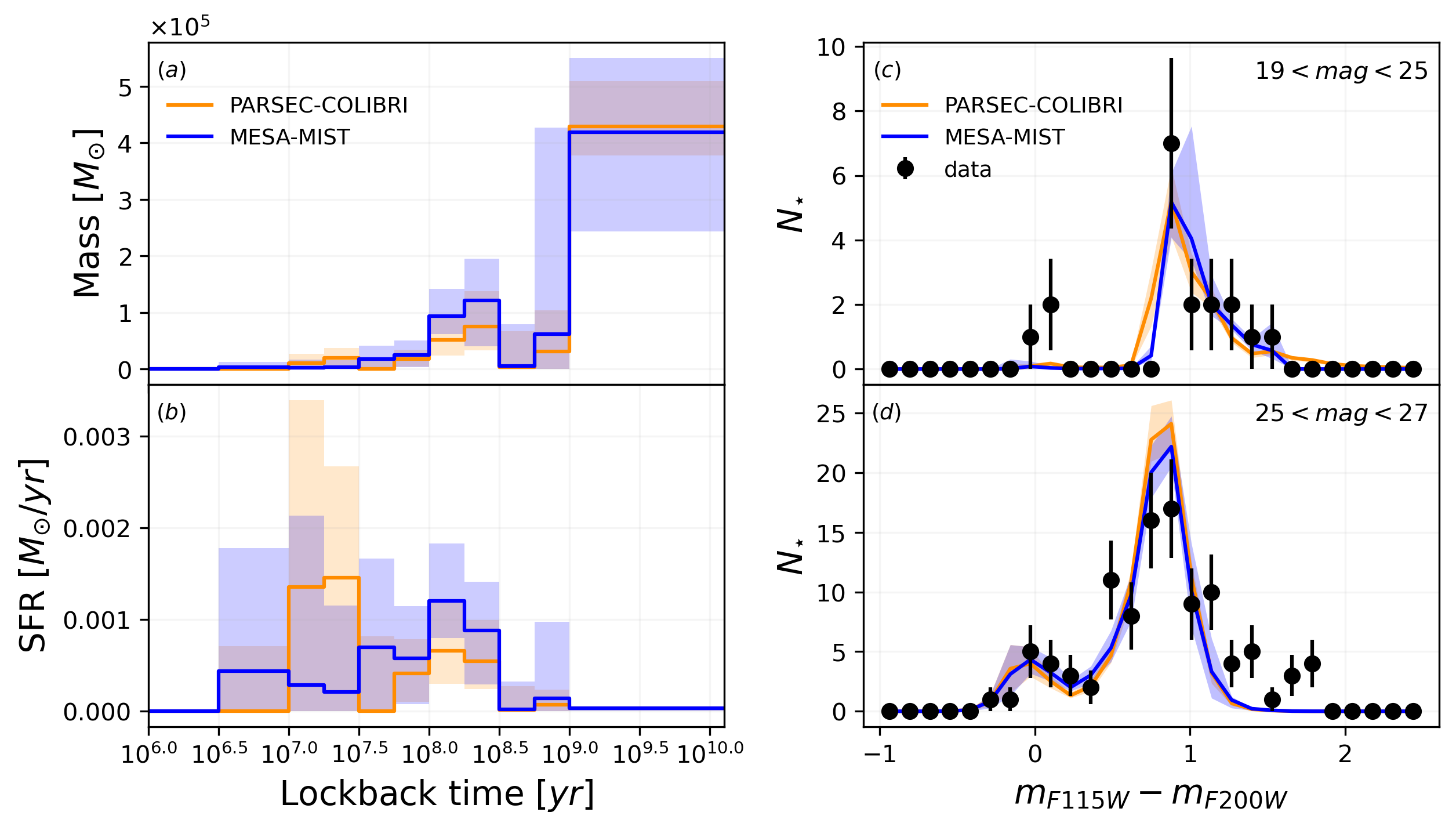}
     \includegraphics[width=17cm]{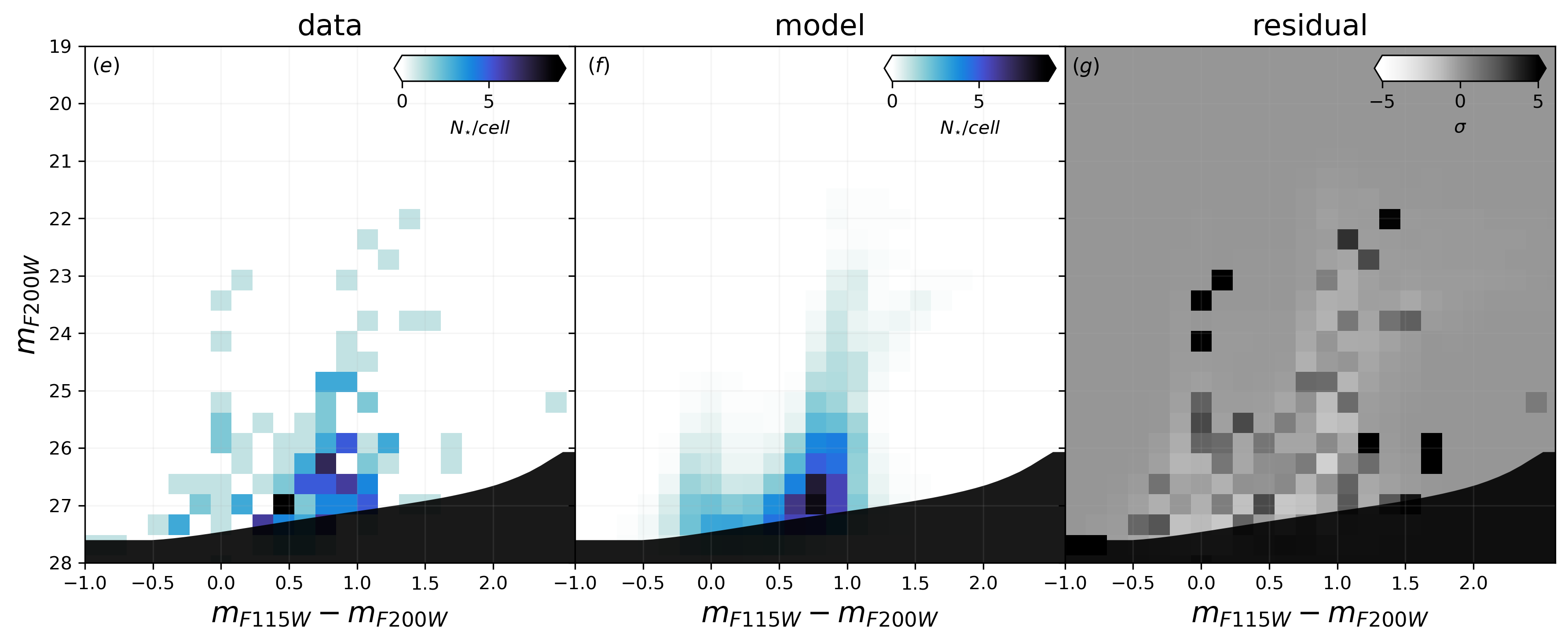}
     \caption{Same as Fig. \ref{fig:region1_results}, but for region 4 of the main body of I Zw 18.}
     \label{fig:region4_results}
\end{figure*}

Region 4 covers the outskirt part of the main body of I Zw 18, and because of its relatively low crowding, is our most complete catalog, reaching the $50\%$ completeness limit around $\sim 27.5$ (see Fig. \ref{AST_regions_completeness}). On the other hand, this region is poorly populated compared to the others, with $\sim200$ detected stars. The recovered stellar mass and SFH here, shown in Fig. \ref{fig:region4_results} (Panels (a) and (b)), display a trend similar to region 3, though with greater uncertainty due to the low number of stars that makes the fitting routine more sensitive to random statistical fluctuations. Again, both PARSEC-COLIBRI and MIST models are in good agreement within the uncertainties. The \texttt{SFERA} 2.0 best model is in very good agreement with the data, as can be appreciated by both the color functions (panels (c) and (d)) and CMD comparison (panels (e)-(g)).    
\subsection{Component C}
\begin{figure*}
\centering
     \includegraphics[width=17cm]{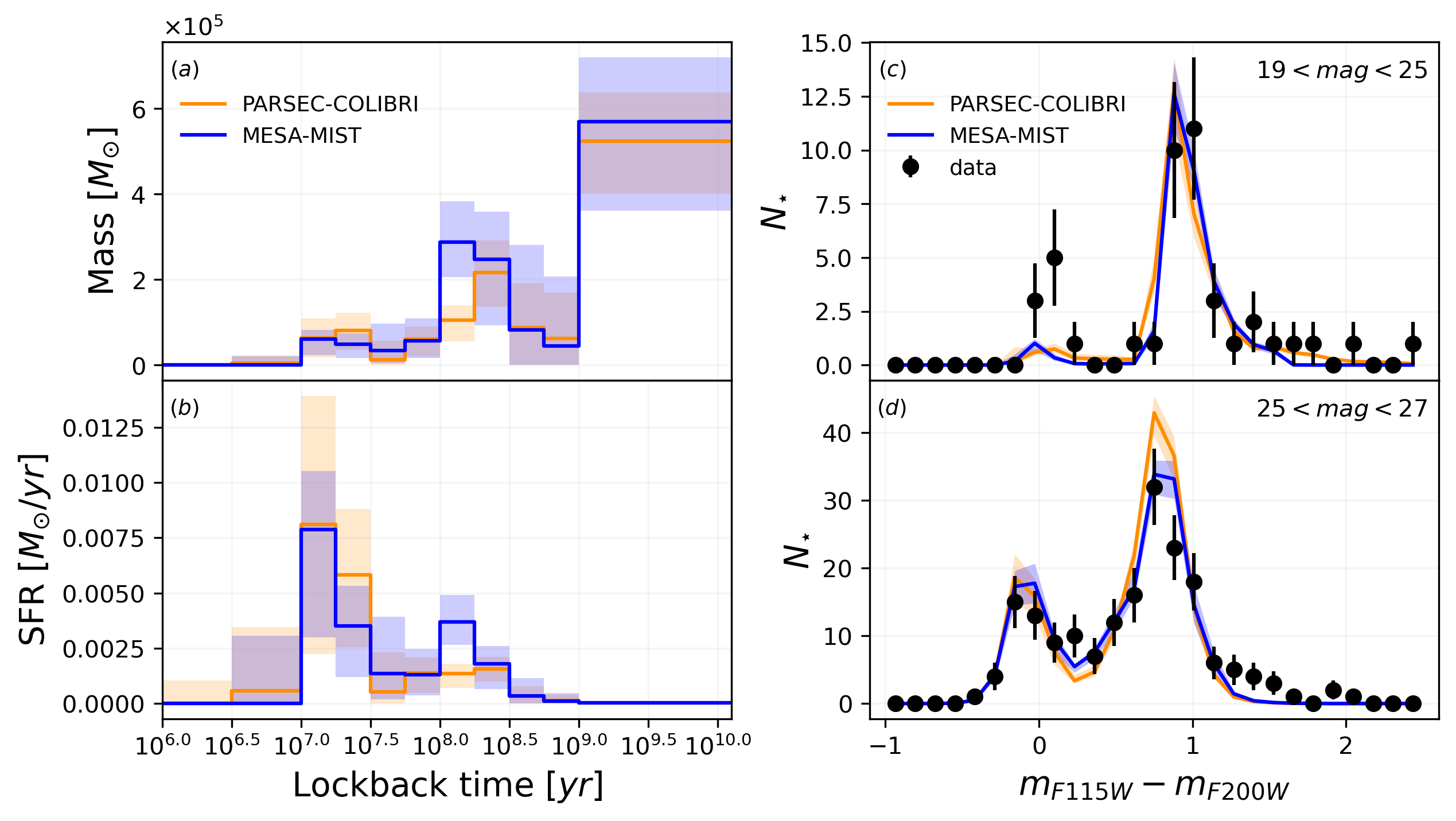}
     \includegraphics[width=17cm]{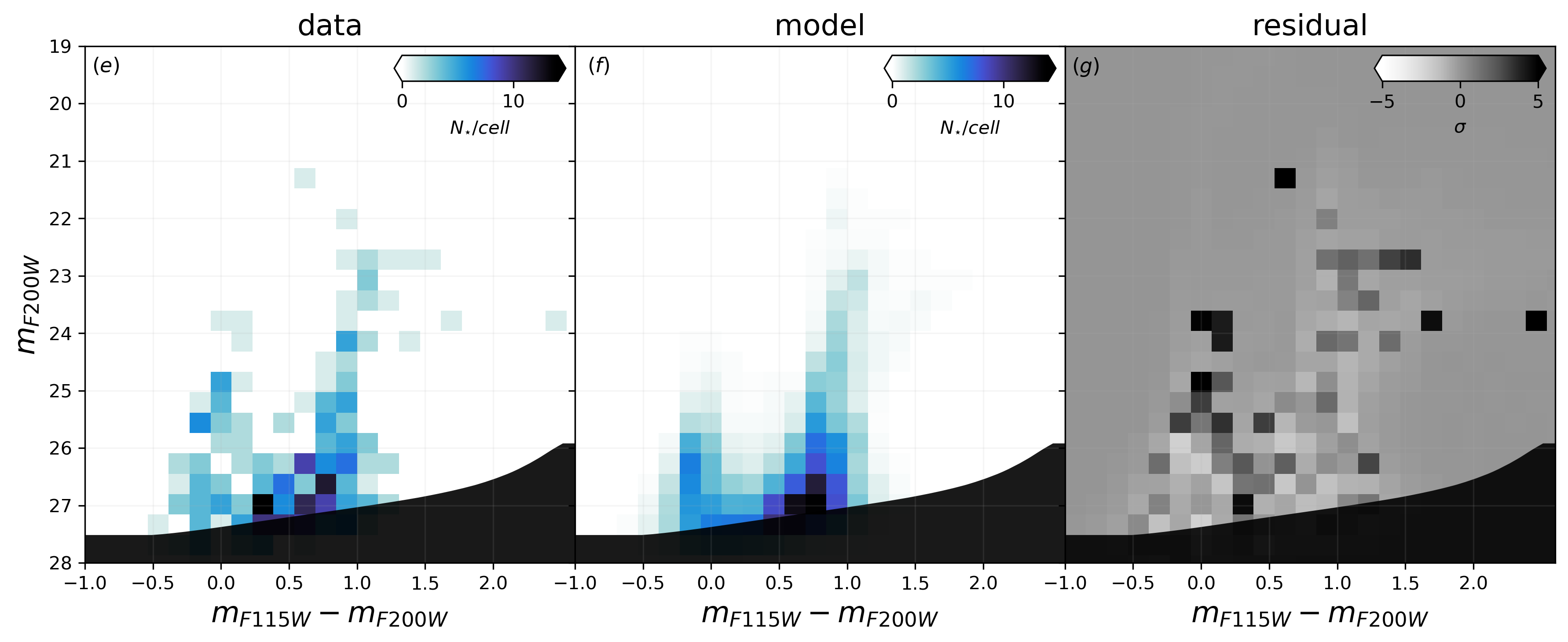}
     \caption{\texttt{SFERA} 2.0 results for I Zw 18 companion (i.e., Component C). See Fig. \ref{fig:region1_results} for a more detailed description of each panel.}
     \label{fig:componentC_results}
\end{figure*}

The CMD of I Zw 18's Component C appears to be quite similar to that of region 3 and 4 in the main body of the galaxy (see right panel of Fig. \ref{I Zw 18_CMD}). In this region, the photometry is as deep as in region 4, allowing us to reach RGB stars $\sim 1$ mag above our fiducial $50\%$ completeness limit, thus effectively sampling the SF activity of this region for epochs older than 1 Gyr. As for region 3 and 4, we adopt a single time bin from $1$ to $14$ Gyr to account for the poor time resolution provided by RGB stars. The results of our synthetic CMD fitting analysis are shown in Fig. \ref{fig:componentC_results}.
The recovered SFH is remarkably similar to the one of region 3, but with lower star-formation rates on average. It shows a very slow start at epochs older than $1$ Gyr (SFR $\sim 10^{-5} \, M_{\sun} \, yr^{-1}$), and then an increasing activity in recent times, characterized by two peaks of SF around $10$ and $100$ Myr ago. Also the recovered stellar mass follows nicely the trend of region 3 and 4. Again, both stellar evolutionary model predictions are in good agreement within errors. The best-model CMD is in excellent agreement with the data, being able to reproduce both the blue and red ``plume'', while accounting for the population of old RGB stars.   

\section{Discussion}\label{sec:Discussion}
Building upon previous I Zw 18 optical CMD studies performed with HST \citep{Hunter1995,Dufour1996HST,Izotov2004,Aloisi2007,ContrerasRamos2011,Annibali2013}, we presented here a detailed study of I Zw 18 main body and Component C stellar populations and star formation histories. Our new JWST/NIRCam observations \citep[see ][]{Hirschauer2024} allowed us to probe with unprecedented resolution and sensitivity the red and evolved stellar populations within this enigmatic galaxy.

Our analysis of I Zw 18's $m_{F115W} - m_{F200W}$ vs. $m_{F200W}$ CMD (see Fig. \ref{I Zw 18_CMD}), compared with some metal-poor isochrones, revealed three main stellar populations: one younger than $\sim30$ Myr, associated mainly with the NW star-forming region; an intermediate-age population ($\sim 100 - 800$ Myr), associated mainly with the SE star forming region; and a red and faint population, linked to the underlying halo of the galaxy, older than $1$ Gyr and possibly as old as the Universe. We emphasize that, together with the study by \cite{Hirschauer2024}, this is the first time RGB stars have been detected down to approximately 1 magnitude below the RGB-tip in this galaxy, thus confirming once and for all the old nature of this system. The stellar populations in the galaxy have the typical spatial distribution of star-forming BCD galaxies \citep[see e.g.,][]{Annibali2003,Papaderos2012,Sacchi2016,Sacchi2021}, with the younger stars clustered toward the center and close to the inner star-forming regions, while intermediate-age and old stars show a more uniform distribution (see Fig. \ref{fig:spatial distribution}). The absence of detected RGB stars in region 1 and 2 is due to the severe intrinsic crowding of these SF regions, limiting our look-back time to $\sim 0.5$ and $\sim 1$ Gyr respectively. However, we found a significant paucity of old stars in the outskirt part of the galaxy, were the depth of our photometry reaches stars older than $1$ Gyr. In particular, the NW side of the main body shows a lack of old stars compared to the SE side (see right-hand side panel of Fig. \ref{fig:spatial distribution}). These findings seem to suggest that the center of mass of the galaxy (i.e., its host dark matter halo) might be shifted toward the SE side, and that the NW region is only recently building up through the current burst of star-formation. Notably, Component C exhibits a mirrored spatial distribution of stars (see Fig. \ref{fig:spatial distribution}), with its young stars clumped toward I Zw 18's main body. This clustering indicates that a close interaction between I Zw 18 and Component C may have occurred $\sim 30-50$ Myr ago that triggered the most current star forming event in both systems \citep{Kim2017}.

The NW SF region of the galaxy (i.e., region 1), shows clear signs of activity in the last $\sim 100$ Myr, that culminates in a very strong bursts around $10$ Myr ago ($\sim 0.6 \, M_{\sun} \, yr^{-1}$). Moving from the NW to the SE SF region (i.e., region 2), we recover a similar trend, with a dominant peak around $\sim 8 - 30$ Myr ago, even if characterized by a substantially lower SF rate compared to region 1. Region 2 also shows signs of mild SF activity at older epochs. Unfortunately, the intrinsic crowding of these regions, together with the presence of very bright, massive stars, imposes strong limitations on the depth reached by our photometry. This hinders us from placing any constraints on the SFHs of these regions at epochs older than approximately $0.8 - 1$ Gyr. Given the presence of old stars in the outskirts of the galaxy, we see no reason why regions 1 and 2 should not host an underlying older population. However, to assess this open question future deeper observations are needed. Moving outward, the CMD becomes deep enough to detect RGB stars that are older than $1$ Gyr, significantly increasing our look-back time. Unfortunately, RGB stars are poor age indicators \citep[e.g., see][]{Gallart2005,Cignoni2010} due to the strong age-metallicity degeneracy that affects this phase. A younger, metal rich RGB stars population can indeed have the same magnitude and color than an older, metal-poor one \citep{Freedman2020}. Other old features, like the red clump or the horizontal branch are too faint ($M_{F200W} \sim -1$ mag) to be detected even with modern telescopes at I Zw 18's distance. This limitation prevents us from precise age-dating beyond $1$ Gyr. For these reasons we made the conservative choice to adopt a single age bin for the interval $1-14$ Gyr. Nevertheless, we are confident that this limitation does not affect the main conclusions of our analysis. In region 3 and 4 the SF rates generally decrease compared to region 1 and 2. Despite the uncertainties, region 3 and 4 show a similar overall history. Both regions show clear traces of SF activity between $1$ and $14$ Gyr ago that has proceeded continuously but at a very low rate ($\sim10^{-4} \; M_{\sun} \; {yr}^{-1}$). Nevertheless, this old population holds the bulk of the galaxy's stellar mass, with a lower limit of $\sim 2 \times 10^{6} \; M_{\odot}$. In the last billion years, both regions show a gradually increasing SF activity, culminating in a very uncertain peak around $10-100$ Myr ago. 

I Zw 18's Component C displays an evolution very similar to the main body, but characterized on average by a lower SFR. As for region 3 and 4, Component C shows a rising trend in SF activity in the last billion years, with two main bursts around $\sim 10$ and $\sim 100$ Myr ago. This recent activity goes on top of what seems to be a very mild SF activity that has proceeded continuously since older epochs. For its stellar mass locked up in stars older than $1$ Gyr, we infer a lower limit of $\sim 6\times10^{5} \, M_{\sun}$.

It is interesting to notice that I Zw 18 shows a very similar stellar populations composition and SFH compared to other metal-poor BCD galaxies studied in the local Universe, such as IZw36 \citep{Schulte-Ladbeck2001}, UGCA 290 \citep{Crone2002}, NGC 1705 \citep{Annibali2003}, NGC 6789 \citep{Garcia-Benito2012}, DDO 68 \citep{Sacchi2018} and UGC 4483 \citep{Sacchi2021}, among others. In all these systems, stars across all mass ranges and in all major phases of stellar evolution (which are above the detection limits reached by the available photometry) have been found, suggesting active SF throughout the whole Hubble time. These results, combined with ours, pose significant constraints on BCDs' chemical evolution models. In particular, I Zw 18 has been forming stars and therefore synthesizing metals for billions of years, and yet its ISM is still extremely metal-poor \citep{Skillman1993,Izotov1999}. This implies that mechanisms like stellar winds, supernovae feedback and galactic outflow \citep{MacLow1999,Hopkins2012}, together with the possible in fall of pristine HI gas \citep{vanZee1998,Lebouteiller2013} must have played a major role in removing or diluting heavy elements from the galaxy \citep{Martin1996}.     
\section{Summary and conclusions}\label{sec:Summary and Conclusions}
Here we summarize the main results of this paper, where we studied I Zw 18's resolved stellar populations and SFH using new JWST/NIRCam data \citep{Hirschauer2024}. I Zw 18 near-infrared CMD is populated by stars of all ages without any substantial gaps, from very young upper-MS to intermediate-age AGB and old RGB stars. In particular, thanks to NIRCam's exquisite sensitivity we were able to detect a population of RGB stars down to $m_{F200W} \sim 27$, which is 1 mag below the RGB-tip. 

Youngest stars are tightly concentrated in the galaxy center, particularly tracing the NW and SE SF regions. Intermediate-age and old AGB and RGB stars, on the other hand, are more uniformly distributed around the center and in the outskirts of the galaxy. We observe a paucity of old stars on the NW side of the main body compared to the SE side. I Zw 18's companion, Component C, displays a mirrored stellar spatial distribution compared to the main body, with the youngest stars clustered toward its SE side. 

From the SFHs recovered for different regions within the main body and Component C, we found an overall rising trend in the SF activity in the last billion years, with two main bursts of SF around $\sim 10$ and $\sim 100$ Myr ago, on top of what seems to be a mild but continuous SF activity that has lasted for the entire Hubble time with an averaged SFR of $10^{-4}-10^{-5}\, M_{\sun} \, yr^{-1}$. This low SFR is in agreement with the ``slow cooking'' dwarf scenario suggested by \citet{Legrand2000,Legrand2001}.

We find that the stellar mass locked up in stars older than $1$ Gyr in I Zw 18's main body and Component C is approximately $1.5 \times 10^6 , M_{\sun}$ and $6 \times 10^5 , M_{\sun}$, respectively. The synchronized recent bursts of SF in I Zw 18's main body and Component C could have been driven by a recent interaction between the two systems \citep{Kim2017}. This result along with the mild star formation activity older epochs, might indicate that Component C has only recently ($\sim 100$ Myr ago) entered I Zw 18's main body gravitational potential. A definite answer would require a dynamical model of the system.
    
Our results confirm that I Zw 18 is not a truly young galaxy but rather a much older system, characterized by an old underlying stellar halo as already suggested by previous HST studies \citep{Aloisi2007,ContrerasRamos2011,Annibali2013}. Additionally, The galaxy is now experiencing its strongest burst of star formation located in the NW region (SFR $\sim 0.6 \, M_{\sun} \, yr^{-1}$). This burst is likely due to a recent gravitational interaction with Component C. Our SFHs will have a crucial role in the ongoing effort to understand the processes at play behind the very low observed metallicity of I Zw 18 and the evolution of star-forming dwarf galaxies.  

\begin{acknowledgements}
    GB expresses his gratitude to Michele Cignoni and Monica Tosi for their insightful and experienced advice. GB thanks Matteo Correnti for the valuable discussions about DOLPHOT and PSF-photometry. GB also thanks Angela Adamo, Matthew Hayes, Jens Melinder, Arjan Bik, Alex Pedrini, Alice Young and all the other members of the Stockholm University Astronomy department for engaging and helpful conversations. GB and GÖ acknowledges support from the Swedish National Space Agency. OCJ acknowledge support from an STFC Webb fellowship. ASH is supported in part by an STScI Postdoctoral Fellowship. CN acknowledges support from an STFC studentship.
    MM and NH acknowledge support through NASA/JWST grant 80NSSC22K0025 and MM and LL acknowledge support from the NSF through grant 2054178.
    MM, NH, and LL acknowledge that a portion of their research was carried out at the Jet Propulsion Laboratory, California Institute of Technology, under a contract with the National Aeronautics and Space Administration (80NM0018D0004). KJ acknowledge the Swedish National Space Agency (SNSA). The authors wish to thank the anonymous reviewer for their insightful comments and suggestions.
\end{acknowledgements}

%
%
\bibliographystyle{aa} 
\bibliography{references.bib} 

\end{document}